\documentclass[journal]{IEEEtran}
\ifCLASSINFOpdf
  \usepackage[pdftex]{graphicx}
   \graphicspath{{../figures/}{../jpeg/}}
\else
\fi
 
\usepackage[cmex10]{amsmath}
\usepackage{amssymb}
\usepackage{bm}
 
\usepackage{algpseudocode}
\usepackage{algorithm}
\usepackage{array}
\usepackage{url}
\usepackage[noadjust]{cite}

\newtheorem{lemma}{Lemma}

\newtheorem{proposition}{Proposition}
\newtheorem{remark}{Remark}

\usepackage{fancyhdr}
\usepackage{amsfonts}
\usepackage{xfrac}
\usepackage{color}
\usepackage{pgfplots}
\pgfplotsset{compat=newest}           
\pgfplotsset{compat=1.17}
\usepackage{enumerate}
 
\usepackage{cite}
\usepackage{mathrsfs}
\usepackage{subcaption}
\usepackage{stfloats}
\usepackage{makecell}
\usepackage{comment}
\usepackage{afterpage}
\usepackage{tikz}
\usepackage{mathdots}
\usepackage{yhmath}
\usepackage{cancel}
\usepackage{color}
\usepackage{siunitx}
\usepackage{array}
\usepackage{multirow}
\usepackage{textcomp}
\usepackage{gensymb}
\usepackage{tabularx}
\usepackage{booktabs}
\usepackage{graphicx}
\usepackage{xcolor}

\allowdisplaybreaks
\usetikzlibrary{decorations.markings,decorations.pathmorphing}

\begin{document}
\title{Integrated Localization, Mapping, and Communication through VCSEL-Based Light-emitting RIS (LeRIS)}
 
\author{Rashid~Iqbal,~\IEEEmembership{Student Member,~IEEE,}
        Dimitrios~Bozanis,~\IEEEmembership{Graduate Student Member,~IEEE,}\\
        Dimitrios~Tyrovolas,~\IEEEmembership{Member,~IEEE,}
        Sotiris~Ioannidis,
        Christos~K.~Liaskos,~\IEEEmembership{Member,~IEEE,}\\
        Muhammad~Ali~Imran,~\IEEEmembership{Fellow,~IEEE,}
        George~K.~Karagiannidis,~\IEEEmembership{Fellow,~IEEE,}\\
        and~Hanaa~Abumarshoud,~\IEEEmembership{Senior Member,~IEEE}%
\thanks{R. Iqbal, M. Imran and H. Abumarshoud are with the James Watt School of Engineering, University of Glasgow, Glasgow, G12 8QQ, UK (r.iqbal.1@research.gla.ac.uk, Muhammad.Imran@glasgow.ac.uk, Hanaa.Abumarshoud@glasgow.ac.uk).}%
\thanks{D. Bozanis and G. K. Karagiannidis are with the Aristotle University of Thessaloniki, 54124 Thessaloniki, Greece (dimimpoz@auth.gr, geokarag@ece.auth.gr).}%
\thanks{D. Tyrovolas is with the Department of Informatics and Telecommunications, University of the Peloponnese, 22132 Tripoli, Greece and with Dienekes SI IKE Heraklion, Crete, Greece (d.tyrovolas@uop.gr).}
\thanks{S. Ioannidis is with the Department of Electrical and Computer Engineering, Technical University of Chania, Chania, Greece and with Dienekes SI IKE Heraklion, Crete, Greece (sotiris@ece.tuc.gr).}
\thanks{C. K. Liaskos is with the University of Ioannina, Greece, and with the Foundation for Research and Technology Hellas (FORTH), Greece (cliaskos@uoi.gr).}%
\thanks{Part of this work has been presented to IEEE International Conference on Communications (ICC) 2026 \cite{IqbalICC2026}.}
\thanks{The authors would like to acknowledge the European Union's Horizon 2020 research and innovation programme through the 6G-XCEL project (GA 101139194), the Horizon Europe project INTACT (GA 101168438), and the EPSRC project UKRI916, PerceptiFi: Integrated Light Sensing and Communications for Perceptive LiFi.}}

\maketitle

 
\begin{abstract}
Light-emitting reconfigurable intelligent surfaces (LeRISs) have recently emerged as a promising solution for providing the spatial awareness required for reliable millimeter-wave (mmWave) communication in programmable wireless environments (PWEs). However, existing LeRIS designs rely on the diffuse emission of light-emitting diodes, while LiDAR-assisted solutions require dedicated sensing modules that hinder their direct integration into RIS panels. In this paper, a vertical-cavity surface-emitting laser (VCSEL)-based LeRIS framework is developed to jointly support user localization, obstacle-aware mapping, and mmWave communication. {Specifically, the narrow Gaussian beams and the beam scanning of VCSELs are exploited to derive closed-form schemes for the joint recovery of the user position and orientation. The analysis shows that five VCSELs are sufficient for locally unique recovery from received signal strength measurements, while this requirement is reduced to three VCSELs when the pointing directions of the scanning beams are exploited under specific geometric conditions.} Furthermore, the position error bound (PEB) is derived to characterize the achievable localization accuracy, while reflected-signal time-of-arrival measurements are employed to identify obstructed links and enable blockage-resilient LeRIS routing. As a result, the proposed framework achieves {millimeter-level} localization accuracy, reliable obstacle detection, and substantial {average-rate} and minimum-user-rate gains, thereby establishing VCSEL-based LeRISs as a solution for spatially aware and resilient PWEs.
\end{abstract}

{
}

%
\IEEEpeerreviewmaketitle
 
\section{Introduction}
The evolution toward sixth-generation wireless networks is fueled by diverse services and applications, such as holographic telepresence, the metaverse, smart healthcare, and industrial automation, which impose stringent requirements on user access, traffic support, bandwidth efficiency, and latency \cite{Holographic,6GMetaverse,6GApplications}. To support these requirements, future systems are expected to exploit the large bandwidth available at millimeter-wave (mmWave) frequencies, whose use is nevertheless challenged by severe propagation loss and high susceptibility to blockage \cite{mmwaveSurvey2017}. To address these limitations, programmable wireless environments (PWEs) have been introduced as a paradigm in which reconfigurable intelligent surfaces (RISs) dynamically reshape wave propagation and render the wireless environment controllable according to the communication objectives \cite{PWELiaskos}. However, the effectiveness of RIS-assisted propagation control depends on accurate knowledge of the user positions and the surrounding obstacles, since inaccurate spatial information can result in unsuitable surface configurations and degraded connectivity. This requirement has motivated the development of multi-functional RISs, which complement wavefront manipulation with additional capabilities such as localization and environmental mapping \cite{MFRIS2025}. As a result, PWEs can acquire the environmental awareness required to identify suitable propagation paths and maintain reliable mmWave communication under dynamic user and blockage conditions.
 
Building upon the need for joint localization, mapping, and communication, multi-functional RISs are envisioned to evolve beyond passive signal control and to support multiple functions that provide further intelligence for PWEs, in addition to their wavefront manipulation role. In this direction, current research primarily pursues this objective by extending the electromagnetic operation of RISs through simultaneous reflection and transmission, together with configuration strategies that allocate its available degrees of freedom among communication, sensing, and localization tasks \cite{MFRIS2025,MFRIS2024}. {For instance, the authors of \cite{Lyu2024CRB} jointly designed the hybrid beamformers and the RIS phase shifts of a mmWave integrated sensing and communication system to minimize the Cram\'{e}r--Rao bound of the target angles under communication rate constraints, characterizing the trade-off between sensing accuracy and communication performance.} Although such approaches preserve a fully electromagnetic implementation, the same surface aperture and configuration resources must support multiple objectives, rendering the achievable communication and sensing performance inherently coupled. Consequently, improving the RIS configuration for one functionality may reduce its effectiveness for another, which motivates lightweight architectural extensions that provide environmental awareness without constraining the wavefront manipulation required for communication.
 
Among the possible architectural extensions, optical sources stand out as candidates that can enhance the environmental awareness of PWEs \cite{Tyrovolas2025Leris}. In more detail, optical sources exhibit propagation characteristics that enable precise extraction of spatial information, since their predominantly line-of-sight (LoS) behavior limits sensitivity to scattering and provides stable references for position and orientation that can be extracted through multiple inherent degrees of freedom, such as intensity or angular features \cite{ghassemlooy2019optical}. At the same time, these measurements can be obtained with low complexity, since spatial parameters evolve slowly, and their operation in a distinct spectral domain ensures that these functions remain isolated from the spectrum manipulated by the RISs. Therefore, optical sources can complement the wavefront manipulation functionality of RISs and enable the joint support of localization, environmental mapping, and communication within the same PWE architecture.
 
\subsection{Literature Review}
{In the RF domain, the authors of \cite{Yang2024BlockageAware} investigated blockage-aware RIS beamforming in mobile mmWave MIMO systems through hypothesis-based blockage detection, adaptive beamwidth control, and beamformer optimization under channel uncertainty. In parallel,} optical systems have been increasingly investigated to provide the environmental awareness required for RIS configuration, with existing approaches primarily relying on LED-based localization or LiDAR-assisted sensing \cite{Tyrovolas2025Leris}. Within the first direction, LEDs have been employed as optical anchors by exploiting their compatibility with existing lighting infrastructures and the dependence of the received optical power on the user geometry. In particular, \cite{boz_loc22} introduced a positioning technique that exploits received signal strength (RSS) measurements from LED anchors through trilateration under channel state information errors, achieving centimeter-level positioning accuracy. The authors of \cite{chaudhary_indoor_2021} subsequently considered tilted transmitters and non-line-of-sight components, while \cite{Wei2023ILAC} developed a federated learning framework that exploits location information to improve communication efficiency. Moreover, closed-form localization expressions for users with random orientations were derived in \cite{boz_loc25}. Alongside LED-based localization, LiDAR measurements have been employed to provide the more detailed environmental information required for mmWave beam management. Specifically, \cite{Klatau2019} showed that LiDAR point-cloud data can support mmWave beam selection by reducing the beam-search overhead without degrading throughput, while \cite{Zecchin2022} demonstrated that near-optimal throughput can be achieved by restricting the search to a small set of candidate beams identified through LiDAR measurements. Finally, \cite{ShunyaoWu2022} exploited LiDAR information for mobile-blockage prediction in real-world mmWave systems, achieving high prediction accuracy with significantly reduced handoff latency. However, even if LEDs provide a low-cost and compact solution that can be integrated into RIS panels, their diffuse emission limits the spatial resolution required for fine-grained environmental mapping. At the same time, LiDAR systems provide accurate three-dimensional information for beam selection and blockage prediction, but their bulky and power-demanding sensing components hinder their direct integration into RIS panels \cite{lidarbulky2022}. Therefore, the two directions provide complementary advantages, while neither simultaneously offers the spatial resolution, compactness, and integrability required to embed localization and mapping capabilities directly into RISs.
 
Building on these observations, vertical cavity surface-emitting lasers (VCSELs) have emerged as compact optical sources that combine precise beam control with fast modulation and low-power operation. Specifically, their highly directional Gaussian-profile beams provide fine angular resolution for accurate localization, while the availability of multiple optical modes offers additional diversity that can be exploited to enhance localization and mapping capabilities in PWEs. In this direction, the authors of \cite{Zeng2021_VCSEL_BeamActivation} proposed a VCSEL array system with novel beam activation schemes that maintained connectivity under random user orientation, while analyzing the tradeoff between divergence and throughput. In addition, \cite{R1} presented a VCSEL-based LiFi approach where a deep neural network estimates user position and orientation jointly from received signals, achieving lower localization error compared to LED-based systems. Finally, \cite{sarbazi} investigated ultra-dense VCSEL array access architectures that supported multi-gigabit transmission through optimized divergence and number of beams, while scanning VCSEL emitters have also been realized through micro-actuated optics in \cite{Hedsten2008}, thereby further reinforcing the potential of VCSELs for environmental mapping. To this end, VCSELs combine compactness, multimode diversity, scalable integration, and dynamic beam-steering mechanisms, which positions them as promising components for realizing multi-functional RISs that jointly support localization, mapping, and communications.

\subsection{Motivation \& Contribution}
The integration of optical anchors into RIS panels has recently led to the concept of light-emitting RISs (LeRISs), where the acquired position information is employed to configure the surface and improve the {achievable rate} of the assisted communication system \cite{Tyrovolas2025Leris,R2}. Nevertheless, existing LeRIS architectures rely on LED sources, whose diffuse emission supports broad coverage but limits the spatial resolution required to identify obstacles and determine whether the corresponding RIS-assisted links remain unobstructed \cite{R2}. As a result, the localization functionality provided by current LeRIS designs cannot be directly extended to the fine-grained environmental mapping required for blockage-aware mmWave communication. Addressing this limitation requires compact optical sources whose emission characteristics can provide both accurate geometric measurements and sufficiently directional spatial observations, while preserving compatibility with the RIS architecture. In this direction, VCSELs provide the required characteristics through their narrow Gaussian beams, multimode operation, compact form factor, and beam-steering capability. In particular, {the distance-dependent Gaussian profiles and the known pointing directions of the scanning beams provide independent measurements for joint position and orientation recovery}, while the narrow and steerable beams can be used to probe the surrounding environment and identify potential blockages. However, existing VCSEL-based studies have considered optical communication, localization, and beam scanning as separate functionalities, without establishing how the resulting spatial information can be acquired analytically and subsequently employed for blockage-aware RIS configuration. Therefore, to the best of the authors' knowledge, no existing work has integrated VCSELs into a LeRIS architecture and jointly developed localization, environmental mapping, and mmWave communication mechanisms within a unified PWE framework.

Motivated by the above, this paper introduces a VCSEL-based LeRIS-assisted PWE architecture that jointly supports localization, mapping, and mmWave communication. The main contributions are summarized as follows:
\begin{itemize}
    \item We introduce a VCSEL-based LeRIS architecture in which VCSELs are integrated along the perimeter of each RIS panel and operate as structured optical anchors for user localization and environmental sensing. 
    \item We derive closed-form localization and orientation estimation schemes that exploit the Gaussian beam profile and {the beam-scanning directionality} of VCSELs, while the Cram\'{e}r--Rao lower bound (CRLB) is also derived for the joint position and orientation state, from which the corresponding position error bound (PEB) is obtained to characterize the achievable localization accuracy.
   \item We develop a VCSEL-based mapping mechanism in which reflected-signal time-of-arrival measurements are associated with the known beam directions to identify obstacle locations and determine whether the candidate LeRIS--UE and LeRIS--LeRIS links are unobstructed. Based on the resulting link states, blockage-resilient LeRIS routes are selected, and a TDMA-based resource-allocation problem is formulated to maximize the {minimum user rate}.
    \item Through numerical simulations, we show that the proposed localization estimator attains the corresponding PEB and provides millimeter-level positioning accuracy when sufficient VCSEL coverage is available. The results also demonstrate reliable obstacle identification under different blockage conditions and show that the route diversity provided by cooperating VCSEL-based LeRIS panels substantially improves the {average rate} and minimum user rate compared with {deployments with fewer LeRIS panels, under both ideal and realistic conditions}.
\end{itemize}

\subsection{Structure of Paper}
 
The rest of the paper is organized as follows: Section II introduces the considered system model, including the VCSEL-based LeRIS architecture and the associated optical and mmWave channel formulations. Section III develops the proposed framework for joint localization, mapping, and communication, while Section IV provides the simulation results, highlighting the accuracy, robustness, and {achievable rate} of the proposed approach under varying conditions. Finally, Section V concludes the paper.
 
\section{System Model\protect\label{sec:II}}
 
\begin{figure}
\centering
\includegraphics[width=0.9\linewidth]{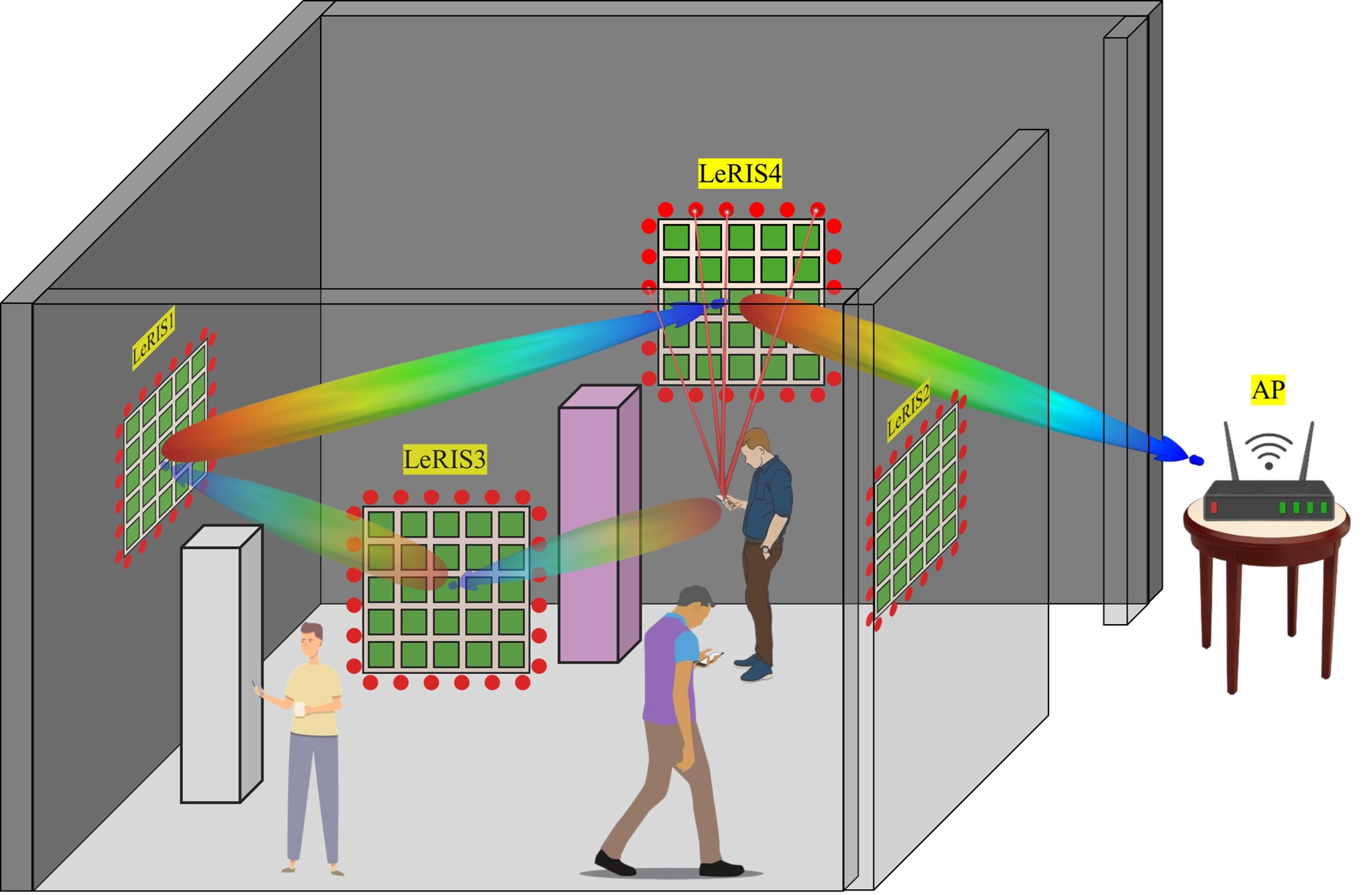}
\caption{Programmable wireless environment with VCSEL-based LeRIS panels.}
\label{fig:system_model}
\end{figure}

We consider a PWE deployed within an indoor space, as shown in Fig.~\ref{fig:system_model}, where a mmWave access point (AP), equipped with a directional antenna and located outside the room, aims to communicate with multiple users (UEs) operating within the enclosed environment. Due to the inherent susceptibility of high-frequency signals to blockage, combined with the presence of obstructing walls and randomly positioned objects, direct LoS transmission from the AP to the UEs becomes infeasible. To overcome these limitations, the PWE leverages four LeRISs, each deployed at the center of a distinct wall \cite{R2}. In more detail, each LeRIS integrates a programmable array of passive reflecting elements, along with optical emitters designed to enable spatially structured illumination for indoor localization and sensing. As a result, the proposed LeRIS-based PWE architecture transforms a static physical space into an adaptive signal manipulation medium, where both electromagnetic reflection and optical transmission are jointly orchestrated to enable dynamic routing, spatial awareness, and resilient wireless access. 
 
To support coordinated access among multiple UEs sharing the same frequency band, the system adopts a time-division multiple access (TDMA) protocol, assigning each time slot to a single user. At the beginning of each slot, {the three-dimensional position and orientation of each UE are available, since they are estimated, concurrently with the mmWave transmission, through the optical signals emitted by} VCSELs integrated along the perimeter of the LeRIS panels. {These VCSELs produce narrow beams with Gaussian profiles and scan assigned angular sectors in azimuth and elevation. The corresponding beam-pointing directions follow a known scanning schedule, so that the angular settings associated with each received-power maximum are available for localization.} Moreover, to facilitate beam identification, each VCSEL transmits at a distinct frequency tone within the infrared spectrum, allowing the UE to spectrally isolate the received signals, infer its position { and orientation}, and report the estimate to the AP over the control channel. In parallel with UE localization, each VCSEL-based LeRIS supports environmental sensing by employing co-located PDs to measure {the ToA of the reflected optical signals}, exploiting the directional nature of the beams and their spatial deployment to detect meaningful reflections from nearby surfaces. By combining the known emission characteristics of the VCSELs with the measured reflections, the PWE constructs a spatial map that identifies existing objects and obstructed LeRIS-UE links. This information, together with the UE's estimated position, is then used to determine viable LoS connections and configure the selected LeRIS panels. As a result, the PWE dynamically adapts its signal routing strategy in real time, maintaining robust and high-throughput multi-user communication in the presence of environmental dynamics and propagation constraints. {It should be noted that the VCSEL scanning, the tone-based identification, and the acquisition of the reflected signals are performed in the optical domain, concurrently with the mmWave transmission, and, thus, they do not occupy its time-frequency resources, while the report of each UE comprises only the five parameters of its estimated position and orientation, whose size does not depend on the number of reflecting elements or on the number of candidate routes.}
 
\subsection{{VCSEL Channel Modeling}}
 
To support high-throughput communication under dynamic indoor conditions, the PWE leverages the deployed LeRISs to form a mmWave signal route toward each selected UE, by utilizing the transmitted optical signals of the VCSELs to extract its location. Unlike conventional diffuse light emitters, such as LEDs, which exhibit broad angular dispersion and lower spatial coherence, VCSELs generate narrow, low-divergence beams with Gaussian spatial intensity profiles that can be tightly controlled and directionally steered, thus being particularly suitable for precise localization.
 
To characterize the optical channel between the $i$-th VCSEL and UE $k$, we consider the emitted intensity profile $I_{i,k}$, which is assumed to follow a Gaussian distribution and is equal to~\cite{R1}
\begin{small}
 \begin{equation}
I_{i,k}(r_0, d_0) = \frac{2P_{t,i}}{\pi w^2(d_0)} \exp\left(-\frac{2r_0^2}{w^2(d_0)}\right),
\label{eq:gaussian_intensity}
\end{equation}   
\end{small}where $P_{t,i}$ is the optical power emitted by the $i$-th VCSEL, $r_0$ is the radial distance from the beam center on the receiver plane, $d_0$ is the axial distance between the VCSEL and the UE, and $w(d_0)$ is the beam spot radius at distance $d_0$, defined as
\begin{small}
\begin{equation}
w(d_0)=w_0 \sqrt{1+\left(\frac{d_0}{z_{\mathrm R}}\right)^{2}}.\label{eq:beam_spot}
\end{equation}    
\end{small}
Here, $w_0$ is the beam waist radius at the VCSEL aperture, defined as $w_0=\frac{\lambda_o}{\pi \theta_{\mathrm{div}}}$, where $\theta_{\mathrm{div}}$ is the divergence angle of the output optical beam from the $i$-th VCSEL, $\lambda_o$ is the wavelength of the VCSEL channel, $z_{\mathrm R}=\frac{\pi w_0^{2}}{\lambda_o}$ represents the Rayleigh range, and $\phi$ is the angle of irradiance with respect to the receiver direction. As a result, the angular intensity distribution at the $k$-th UE is given as
\begin{small}
\begin{equation}
I_{i,k}(d, \phi) = \frac{2P_{t,i}}{\pi w^2(d_{i,k} \cos \phi)} \exp\left(-\frac{2d_{i,k}^2 \sin^2 \phi}{w^2(d_{i,k} \cos \phi)}\right),
\label{eq:angled_intensity}
\end{equation}    
\end{small}
where $d_{i,k}$ is the distance between the $i$-th VCSEL and the $k$-th UE. Based on this intensity, the received LoS optical power at the PD of the $k$-th UE is given by
\begin{small}
\begin{equation}
P_{\mathrm{LoS},i,k} = I_{i,k}(d, \phi) A_{\mathrm{PD}} \cos \psi_{i,k} \, \mathrm{rect}\left(\frac{\psi_{i,k}}{{2\Psi}}\right),
\label{eq:received_power}
\end{equation}    
\end{small}
where $A_{\mathrm{PD}}$ is the active area of the PD, $\psi_{i,k}$ is the angle of incidence, $\Psi$ is the receiver's half-angle field of view (FoV), and $\mathrm{rect}(\cdot)$ is the rectangular function which ensures that optical signals are discarded if the incidence angle exceeds the FoV. However, the RSS measured at the PD is also affected by
receiver noise. To maintain the observation model in the
optical-power domain, the RSS associated with the $i$-th
VCSEL at UE $k$ is modeled as
\begin{small}
 \begin{equation}
P_{r,i,k}
=
P_{\mathrm{LoS},i,k}
+
\nu_{i,k},
\qquad
\nu_{i,k}
\sim
\mathcal N
\left(
0,P_{n,i,k}^{2}
\right),
\label{eq:received_optical_power}
\end{equation}   
\end{small}where $P_{n,i,k}$ denotes the standard deviation of the equivalent optical noise, expressed in watts.
 
Let $B_o$ denote the effective noise-equivalent bandwidth of the optical receiver~\cite{Kazemi}{, which, since each VCSEL is identified through its distinct frequency tone, is determined by the narrowband filter that isolates this tone}. The equivalent
optical-noise variance is then given by
\begin{small}
\begin{equation}
P_{n,i,k}^{2}
=
\frac{B_oS_{i,k}}{R_{\mathrm{PD}}^{2}}=\frac{\sigma_{I,i,k}^{2}}{R_{\mathrm{PD}}^{2}},
\label{eq:equivalent_optical_noise_variance}
\end{equation}    
\end{small}where $S_{i,k}$ is the single-sided electrical current-noise
power spectral density at the PD output, expressed in
$\mathrm{A}^{2}/\mathrm{Hz}$. It is given by
\begin{small}
\begin{equation}
S_{i,k}
=
A_K
+
2qR_{\mathrm{PD}}P_{\mathrm{LoS},i,k}
+
\mathrm{RIN}
\left(
R_{\mathrm{PD}}P_{\mathrm{LoS},i,k}
\right)^2,
\label{eq:electrical_noise_psd}
\end{equation}    
\end{small}where
$A_K=4k_BT F_n/R_L$ denotes the thermal-noise power spectral density. Moreover, $k_B$ is the Boltzmann constant, $T$ is the absolute temperature, $R_L$ is the load resistance, $F_n$ is the noise figure of the preamplifier, $q$ is the elementary charge, $R_{\mathrm{PD}}$ denotes the
PD responsivity, and $\mathrm{RIN}$ denotes the relative-intensity-noise power spectral density. To this end, the optical channel model captures the physical beam propagation, receiver orientation, and power conversion dynamics, which together enable reliable localization and sensing for the PWE.
 
\subsection{{mmWave Channel Modeling}}
 
{Based on the acquired location and sensing information, the AP selects a cascaded route $\pi$ of $L_\pi$ participating LeRISs that connects the AP to UE $k$, where each LeRIS along this path is configured to steer the incident mmWave signal toward the next LeRIS, while the final LeRIS steers energy toward the UE according to its estimated angular location $(\hat{\theta}_{r,k},\hat{\phi}_{r,k})$.} For this purpose, each LeRIS panel, composed of $M \times N$ square reflecting elements with side length $D$, shapes the outgoing wavefront based on the intended direction. Thus, the effective gain of each LeRIS produced in the target direction $(\theta, \phi)$ is given by
\begin{small}
\begin{equation}
    G \left(\theta, \phi \right) = {\eta_{\mathrm{eff}}}  \dfrac{4 \pi \lvert F\left(\theta, \phi \right)\rvert ^2}{\int_{0}^{2\pi} \int_{0}^{\frac{\pi}{2}}\lvert F\left(\theta, \phi \right)\rvert ^2 \mathrm{sin}\left(\theta\right) d\theta d\phi},
\label{eq:leris_gain}
\end{equation}    
\end{small}

where $\eta_{\mathrm{eff}}$ denotes the efficiency factor multiplying the ideal array directivity \cite{cui}. Moreover, $F(\theta, \phi)$ is the far-field radiation pattern generated by the LeRIS array, which is given as 
\begin{small}
 \begin{equation}
F(\theta, \phi) = \sum_{m=1}^M \sum_{n=1}^N e^{j\left(k_0 \zeta_{mn}(\theta,\phi) + \omega_{mn} + \Phi_{mn}\right)},
\end{equation}   
\end{small}where $k_0 = \frac{2\pi}{\lambda_m}$ is the wavenumber and $\lambda_m$ is the wavelength of the mmWave channel, and the term $\zeta_{mn}(\theta, \phi)$ expresses the angular-dependent phase shift associated with the array's geometry and can be expressed as
\begin{small}
\begin{equation}
\begin{split}
\zeta_{mn}(\theta, \phi) = D \sin(\theta) \left[ \left(m - \tfrac{1}{2} \right) \cos(\phi) + \left(n - \tfrac{1}{2} \right) \sin(\phi) \right] \\
+ (x_T - x_{R}) \sin(\theta) \cos(\phi)
+ (y_T - y_{R}) \sin(\theta) \sin(\phi) \\
+ (z_T - z_{R}) \cos(\theta),
\end{split}
\end{equation}
\end{small}{where $\mathbf p_T=(x_T,y_T,z_T)^T$ is the position of the transmitting node, i.e., the AP or the preceding LeRIS of the route, and $\mathbf p_R=(x_R,y_R,z_R)^T$ is the center of the considered LeRIS. Moreover, $\mathbf e_1$ and $\mathbf e_2$ are the unit vectors that span its surface along the horizontal and vertical directions, respectively, and $\mathbf n_R$ is its unit normal, from which $\theta$ is measured, while $\phi$ is measured from $\mathbf e_1$.} Additionally, the phase offset $\omega_{mn}$, corresponding to the path delay from {the transmitting node} to the $(m,n)$-th reflecting element, is given by
\begin{small}
\begin{equation}
\omega_{mn}=k_0\|\mathbf p_T-\mathbf p_{mn}\|,
\label{eq:corrected_incident_phase}
\end{equation}
\end{small}{where $\mathbf p_{mn}=\mathbf p_R+D\left(m-\frac{M+1}{2}\right)\mathbf e_1+D\left(n-\frac{N+1}{2}\right)\mathbf e_2$ denotes the position of the $(m,n)$-th element.} Finally, to steer energy accurately in the direction $(\hat{\theta}_{r,k}, \hat{\phi}_{r,k})$, the phase shift $\Phi_{mn}$ applied by each element is configured as \cite{Abadal}
\begin{small}
\begin{equation}
\begin{split}
\Phi_{mn}& = -k_0 D \left[m \cos(\hat{\phi}_{r,k}) \sin(\hat{\theta}_{r,k}) + n \sin(\hat{\phi}_{r,k}) \sin(\hat{\theta}_{r,k}) \right] \\& - \omega_{mn}.
\end{split}
\end{equation}    
\end{small}
{The RIS configuration above corresponds to the nominal geometry. To model hardware actuation errors, the phase applied by the $(m,n)$-th element of the $\ell$-th participating LeRIS is expressed as \cite{Wang2021VonMises}
\begin{small}
\begin{equation}
\widetilde{\Phi}_{\ell,mn}
=\Phi_{\ell,mn}-\widetilde{\theta}_{{\rm hw},\ell,mn},
\label{eq:hardware_phase_error}
\end{equation}    
\end{small}where the errors $\widetilde{\theta}_{{\rm hw},\ell,mn}$ are independent von Mises random variables with zero circular mean and concentration parameter $\kappa_{\rm hw}$. The actual radiation pattern $\widetilde F_{\ell}(\theta,\phi)$ is obtained by substituting $\widetilde{\Phi}_{\ell,mn}$ for $\Phi_{\ell,mn}$ in the array factor above, and the corresponding directional gain $\widetilde G_{\ell}$ follows from the same gain normalization. In addition, a panel-position offset is represented by $\mathbf p_{\ell}^{\rm act}=\mathbf p_{\ell}^{\rm nom}+\Delta_o\mathbf v_{\ell}$, where $\mathbf v_{\ell}$ is a unit vector parallel to its mounting wall. The nominal positions determine the configured steering phases, whereas the actual positions determine the propagation distances, angles, and resulting received signal. These impairments are introduced only when evaluating the performance of an already selected route, and not when deciding which route to select.}

{When steering toward another LeRIS panel, the nominal position of the subsequent node is known and is used to set the phase profile. For ideal phase control and nominal geometry, the corresponding maximum gain is}
\begin{small}
 \begin{equation}
    G_{\mathrm{max}} {\approx} \eta_{\mathrm{eff}}{\frac{4\pi MND^2}{\lambda_m^2}},
\end{equation}   
\end{small}
{i.e., the directivity of the LeRIS aperture scaled by $\eta_{\mathrm{eff}}$, which is equal to $\eta_{\mathrm{eff}}\pi MN$ for $D=\lambda_m/2$.}
{Under phase-actuation errors or panel offsets, the intermediate-panel gain is instead evaluated from its actual radiation pattern in the direction of the subsequent LeRIS. Thus, the cumulative gain of the first $L_\pi-1$ participating panels is}
\begin{small}
\begin{equation}
G_{\mathrm{cas}}
=\prod_{\ell=1}^{L_\pi-1}
\left(\widetilde A_{\mathrm{eff}}\widetilde G_{\ell}\right),
\label{eq:cascaded_gain}
\end{equation}    
\end{small}
where
\begin{small}
\begin{equation}
A_{\mathrm{eff}} = \frac{MN \lambda_m^2}{4\pi},
\end{equation}    
\end{small}denotes the effective aperture of each LeRIS panel \cite{Abadal}{, and, thus, its normalized counterpart $\widetilde A_{\mathrm{eff}}=4\pi A_{\mathrm{eff}}/\lambda_m^2=MN$ represents the dimensionless gain with which each LeRIS collects the impinging power}. However, for the last participating LeRIS in the route, the signal will be steered toward the estimated UE direction $(\hat{\theta}_{r,k}, \hat{\phi}_{r,k})$, derived from the proposed optical localization system. Specifically, due to potential angular mismatch between the estimated and true UE direction $(\theta_{u,k}, \phi_{u,k})$, the effective beam alignment may degrade, resulting in a reduced directional gain \cite{R2}. As a result, the achievable gain by the final LeRIS in the true UE direction is expressed as
\begin{small}
\begin{equation}
{G_{L_\pi}}(\theta_{u,k}, \phi_{u,k}) = \eta_{\mathrm{eff}} \cdot \frac{4\pi \left|{\widetilde F_{L_\pi}}(\theta_{u,k}, \phi_{u,k})\right|^2}{\int_0^{2\pi} \int_0^{{\pi/2}} \left|{\widetilde F_{L_\pi}}(\theta, \phi)\right|^2 \sin \left(\theta\right) d\theta d\phi}.
\end{equation}    
\end{small}Consequently, the total directional gain accumulated along the entire route is expressed as
\begin{small}
\begin{equation}
{G_{\pi,k}} = G_{\mathrm{cas}} \cdot {\widetilde A_{\mathrm{eff}}} \cdot {G_{L_\pi}}(\theta_{u,k}, \phi_{u,k}),
\end{equation}    
\end{small}where {the factor $\widetilde A_{\mathrm{eff}}$ of the final LeRIS accounts for its normalized effective aperture}. {It should be noted that, since $\eta_{\mathrm{eff}}$ multiplies the gain of each participating LeRIS in \eqref{eq:leris_gain}, $G_{\pi,k}$ is scaled by $\eta_{\mathrm{eff}}^{L_\pi}$, and, thus, the efficiency loss accumulates over the $L_\pi$ reflections of route $\pi$.} {As such, the baseband signal received by UE $k$ over route $\pi$ can be written as
\begin{small}
\begin{equation}
y_k=\sqrt{l_pG_tG_rP_tG_{\pi,k}}\,x+w_k,
\end{equation}    
\end{small}where $x$ is the transmitted symbol with unit energy, $G_t$ and $G_r$ are the antenna gains of the AP and UE antennas, respectively, $P_t$ is the transmit power, and $w_k$ denotes the additive white Gaussian noise at the UE with variance $\sigma_k^2$. Moreover, for a route $\pi$ that consists of $L_\pi$ LeRISs, $l_p$ denotes the total path loss over its $L_\pi+1$ segments and is given by \cite{Cascaded}
\begin{small}
\begin{equation}
l_p=\prod_{s=1}^{L_\pi+1}C_0\left(\frac{d_{r,s}}{d_{r,0}}\right)^{-n_s},
\label{eq:path_loss}
\end{equation}    
\end{small}where $d_{r,s}$ represents the distance of the $s$-th segment, $n_s$ is the path loss exponent of the $s$-th segment, and $C_0=\frac{\lambda_m^2}{(4\pi d_{r,0})^2}$ is the free space path loss at the reference distance $d_{r,0}$. Accordingly, by defining the transmit signal-to-noise ratio as $\gamma_t=P_t/\sigma_k^2$, which does not include the propagation losses of the route, since these are captured by $l_p$ in \eqref{eq:path_loss}, where $\sigma_k^2=-174~\text{dBm/Hz}+10\log_{10}B+\mathrm{NF}$ denotes the noise power at the UE receiver, $B$ denotes the system bandwidth, and NF is the noise figure of the receiver, the rate of UE $k$ over an unobstructed route $\pi$ is given by
\begin{small}
\begin{equation}
R_{\pi,k}=B\log_2\left(1+l_pG_tG_r\gamma_tG_{\pi,k}\right),
\label{eq:spectral_efficiency}
\end{equation}    
\end{small}}where \eqref{eq:spectral_efficiency} captures the effects of
the cascaded path loss, antenna gains, and LeRIS beam-steering
accuracy on the end-to-end {rate}.
 
\section{Localization, Mapping and Communication through VCSEL-based LeRIS}
This section develops a unified framework that leverages the VCSEL-based LeRIS infrastructure to jointly support user localization, environmental sensing, and {multi-user} mmWave communication. First, we derive a closed-form localization method that exploits the spatial intensity profile of the VCSEL beams to estimate each UE position. Next, we present a sensing scheme that employs the same VCSEL infrastructure to detect obstacles and assess the feasibility of LoS links between the UEs and the deployed LeRISs. Finally, a time-resource allocation problem is formulated and solved to promote fairness by maximizing the {minimum user rate} across all UEs.

\subsection{VCSEL-based Localization} 
To determine the position of the $k$-th UE within the PWE, the RSS measurements from the VCSELs need to be utilized, since the attenuation of the optical beam intensity with distance provides geometric information for localization. However, the highly directional emission of VCSELs improves spatial resolution, but limits coverage, making user detection and VCSEL placement dependent on whether the receiver lies within the narrow beam. Specifically, as it can be seen in \eqref{eq:received_power}, a PD can only receive power from a VCSEL if it lies within the narrow beam's footprint, making it essential to ensure that enough VCSELs are deployed for reliable estimation of both the UE's position and orientation. This motivates the derivation of the minimum number of VCSELs that must be integrated onto a LeRIS panel to ensure that a UE within the PWE can be {localized}, regardless of its position. {For brevity, the UE index $k$ is omitted in Lemma~\ref{lemma1} and Proposition~\ref{prop1}.} In this direction, the following lemma establishes the minimum number of VCSELs required {for the local identifiability of the UE position and orientation}, based on RSS values.
 
\begin{lemma} \label{lemma1}
The position and the orientation of the UE {are locally identifiable} from 5 VCSELs if the RSS from the VCSELs within the PD FoV is described by $P_i = \beta_i(d_i)\,{\mathbf{n}^T\mathbf{u}_i}$, where $i=1,\dots,5$, $\mathbf{s}_i\in\mathbb{R}^3$ denotes the position of the $i$-th VCSEL, $\mathbf{r}\in\mathbb{R}^3$ is the position of the UE's PD, $\mathbf{n}\in\mathbb{R}^3$ is the unit normal vector of the PD surface, $d_i=\|\mathbf{s}_i-\mathbf{r}\|$ is the distance between the $i$-th VCSEL and the UE, $\mathbf{u}_i=\tfrac{\mathbf{s}_i-\mathbf{r}}{d_i}$ is the unit vector pointing from the PD toward the $i$-th VCSEL, and $\beta_i(d)=\tfrac{2 P_{t,i} A_{\mathrm{PD}}}{\pi w^2 (d)}$, provided that the VCSELs contribute independent RSS values.  
\end{lemma}
\begin{IEEEproof}
Let the $i$-th VCSEL be at $\mathbf{s}_i\in\mathbb{R}^3$ and the PD at $\mathbf{r}\in\mathbb{R}^3$, with unit surface normal $\mathbf{n}$ satisfying $\|\mathbf{n}\|=1$. Considering the definitions of $d_i$ and the unit direction vector $\mathbf{u}_i$, we can express $\cos\psi_i$ in \eqref{eq:received_power} as $\cos\psi_i={\mathbf{n}^T\mathbf{u}_i}$. Moreover, since each VCSEL sweeps its assigned angular segment and the RSS measurement is acquired at the instant of maximum received power, which corresponds to boresight alignment between the beam axis and the PD, it holds that $\phi = 0^{\circ}$. Thus, given that each VCSEL points within the UE's FoV, \eqref{eq:received_power} simplifies to
\begin{small}
\begin{equation}
P_i=\beta_i(d_i)\,{\mathbf{n}^T\mathbf{u}_i}.
\label{modelp}
\end{equation}    
\end{small}As it can be seen, for each $i$-th VCSEL, this relation is linear in $\mathbf{n}$ once $\mathbf{r}$ is fixed, since then $d_i$ and $\mathbf{u}_i$ are known quantities. Hence, each RSS measurement contributes one scalar linear equation in the three components of $\mathbf{n}$. Therefore, considering $N_{\mathrm{V}}$ RSS measurements, we obtain the following system of equations,
\begin{small}
 \begin{equation}
V(\mathbf{r})\mathbf{n}=\mathbf{p},
\label{eq:system}
\end{equation}   
\end{small}
where $\mathbf{p}=[P_1,\dots,P_{N_{\mathrm{V}}}]^{T}$, and $V(\mathbf{r})$ can be expressed as
\begin{small}
\begin{equation}
V(\mathbf{r})\triangleq
\begin{bmatrix}
(\beta_1(d_1)\mathbf{u}_1)^{T}\\[-1mm]
\vdots\\[-1mm]
(\beta_{N_{\mathrm{V}}}(d_{N_{\mathrm{V}}})\mathbf{u}_{N_{\mathrm{V}}})^{T}
\end{bmatrix}\!\in\mathbb{R}^{N_{\mathrm{V}}\times 3}.
\label{eq:stacked}
\end{equation}    
\end{small}
To acquire the minimum number of VCSELs needed for localization, the required number of independent RSS equations for a {locally} unique solution must be identified. In the ideal case of noise-free measurements and perfect VCSEL-PD alignment, the RSS from each VCSEL coincides exactly with \eqref{modelp}. Under this condition, an exact solution of {\eqref{eq:system}} exists whenever the number of independent equations matches the number of unknowns, which directly reveals the minimum number of VCSELs required. In practice, however, the measured powers $P_i$, although dominated by the LoS component, deviate from the ideal model due to additive noise, as shown in \eqref{eq:received_optical_power}, and imperfections of the Gaussian beam approximation. As a result, the measurement constraints do not intersect at a single point, and direct inversion of the system becomes inconsistent. To overcome this, while incorporating all observations in a coherent manner, the localization problem can be reformulated as a constrained least-squares optimization that minimizes the aggregate residual error across all VCSEL measurements, while enforcing the physical condition ${\|\mathbf{n}\|}=1$, leading to 
\begin{small}
\begin{equation}
\min_{\mathbf{r}\in\mathbb{R}^3,{\|\mathbf{n}\|}=1}
\frac{1}{2}{\left\|V(\mathbf{r})\mathbf{n}-\mathbf{p}\right\|_2^2}.
\label{eq:main_problem}
\end{equation}    
\end{small}To address \eqref{eq:main_problem}, it is noted that, while both $\mathbf{r}$ and $\mathbf{n}$ are unknown, the dependence on $\mathbf{n}$ is linear once $\mathbf{r}$ is fixed, which renders the formulation separable and allows $\mathbf{n}$ to be expressed in closed form, so that the optimization reduces to a search over $\mathbf{r}$. Therefore, for a given $\mathbf{r}$, determining $\mathbf{n}$ reduces to the constrained problem
\begin{small}
\begin{equation}
\min_{{\|\mathbf{n}\|}=1}\tfrac{1}{2}{\|V\mathbf{n}-\mathbf{p}\|_2^2},
\label{eq:subproblem}
\end{equation}    
\end{small}To handle the unit-norm constraint, the Lagrangian method is employed, which yields the regularized normal equations,
\begin{small}
\begin{equation}
\big(V^{T} V+\lambda I_3\big)\mathbf{n}=V^{T} \mathbf{p},
\label{eq:reg_normal}
\end{equation}    
\end{small}where the multiplier $\lambda$ must be selected such that ${\|\mathbf{n}\|}=1$. Thus, after some algebraic manipulations, we obtain
\begin{small}
\begin{equation}
\mathbf{n}(\lambda)=\big(V^{T} V+\lambda I_3\big)^{-1}V^{T} \mathbf{p},
\end{equation}    
\end{small}
and
\begin{small}
 \begin{equation}
    \phi(\lambda)={\|\mathbf{n}(\lambda)\|_2^2}-1.
\end{equation}   
\end{small}Since $\phi(\lambda)$ is strictly decreasing on $(-\sigma_{\min}^2,\infty)$, with $\sigma_{\min}$ denoting the smallest singular value of $V$, {while $\phi(\lambda)\to-1$ as $\lambda\to\infty$ and $\phi(\lambda)\to\infty$ as $\lambda\to-\sigma_{\min}^2$ whenever $V^T\mathbf{p}$ has a nonzero component along the right singular vector of $V$ associated with $\sigma_{\min}$, which holds with probability one for noisy RSS values,} there exists a unique $\hat{\lambda}$ satisfying $\phi(\hat{\lambda})=0$, which implies that, for each $\mathbf{r}$, the orientation is uniquely recovered as $\hat{\mathbf{n}}(\mathbf{r})=\mathbf{n}(\hat{\lambda})$, thereby confirming that $\mathbf{n}$ is fully determined once $\mathbf{r}$ is fixed. Thus, the optimization problem \eqref{eq:subproblem} can be rewritten as 
\begin{small}
\begin{equation}
\hat{\mathbf{r}}=\arg\min_{\mathbf{r}\in\mathbb{R}^3}J(\mathbf{r}),
\label{minproblem}
\end{equation}    
\end{small}where $J(\mathbf{r})=\tfrac{1}{2}{\|V(\mathbf{r})\hat{\mathbf{n}}(\mathbf{r})-\mathbf{p}\|_2^2}$, and $\hat{\mathbf{n}}=\hat{\mathbf{n}}(\hat{\mathbf{r}})$. Although $J(\mathbf{r})$ is continuously differentiable, it is not globally convex, because $d_i$ and $\mathbf{u}_i$ depend nonlinearly on $\mathbf{r}$, which may lead to saddle points or local minima under unfavorable VCSEL placement or noise. Nevertheless, since each candidate $\mathbf{r}$ corresponds to a single compatible orientation, \eqref{minproblem} can be solved via the Levenberg–Marquardt method in variable-projection form, as shown in Appendix \ref{App1}. To this end, since the unknown parameters are the three spatial coordinates of $\mathbf{r}$, and the two orientation angles of $\mathbf{n}$ constrained by unit norm, it follows that at least five VCSELs are required to ensure a {locally} unique solution{, which is attained when the VCSELs contribute independent RSS values, i.e., when the Jacobian of the RSS values with respect to these five parameters has full column rank, since, then, by the inverse function theorem, the mapping from the UE state to the RSS values is one-to-one within a neighborhood of the true state}, which concludes the proof.
\end{IEEEproof}
\begin{remark}
    The deployment of VCSELs on LeRIS panels provides a practical advantage for localization, since their positions and orientations can be arranged by design and are therefore known at the receiver, ensuring that the RSS equations arise from a controlled geometry that mitigates unfavorable arrangements, such as nearly collinear or insufficiently separated VCSELs. 
\end{remark}
 
{From Lemma~\ref{lemma1}, it follows that five VCSELs are sufficient for the joint recovery of the UE position and the PD orientation vector from their RSS values. However, the required number of VCSELs is further reduced by exploiting the directional information that the angular sweep provides, since, at the instant of maximum received power, the pointing direction of each VCSEL coincides with its direction toward the UE. As a result, each VCSEL provides two angular observations in addition to its RSS value, thereby enriching the available set of measurements without increasing the number of the utilized VCSELs. In this direction, the following proposition establishes that the number of VCSELs required for unique localization is reduced to three, when the VCSELs are appropriately deployed on the LeRIS panels and satisfy certain geometric conditions.}
\begin{proposition}
\label{prop1}
    The position and orientation of the UE {are uniquely determined from three VCSELs}, provided that the following conditions hold.
    \begin{enumerate}
        \item The VCSELs are placed at non-collinear positions.
        \item {Each VCSEL sweeps its beam in azimuth and elevation, such that its pointing direction at the instant of maximum received power is known.}
        \item The unit vectors from each VCSEL to the UE, defined as $\mathbf{u}_i = \frac{\mathbf{s}_i - \mathbf{r}}{\|\mathbf{s}_i - \mathbf{r}\|}$, with $\quad i=1,2,3,$ are linearly independent.
    \end{enumerate}
\end{proposition}
\begin{IEEEproof}


Let the $i$-th VCSEL be at $\mathbf{s}_i\in\mathbb{R}^3$, and the UE at $\mathbf{r}\in\mathbb{R}^3$, with unit normal $\mathbf{n}\in\mathbb{R}^3$ satisfying ${\|\mathbf{n}\|}=1$. {Following the sweep-based RSS acquisition described in Lemma~\ref{lemma1}, only measurements satisfying $\mathbf{n}^T\mathbf{u}_i\geq\cos\Psi$ are retained. In the absence of noise, the RSS of the $i$-th VCSEL at the instant of maximum received power is given by \eqref{modelp}, where the orientation factor $\mathbf{n}^T\mathbf{u}_i$ does not depend on the pointing direction of the beam. Since the Gaussian beam intensity is maximized when the beam axis passes through the PD, the pointing direction $\mathbf{b}_i$ of the $i$-th VCSEL at this instant, which is known from its sweep, satisfies}
\begin{small}
\begin{equation}
\mathbf{b}_i=
\frac{\mathbf{r}-\mathbf{s}_i}
{\|\mathbf{r}-\mathbf{s}_i\|},
\label{eq:pointing_direction}
\end{equation}
\end{small}
{where the orientation factor $\mathbf{n}^T\mathbf{u}_i$ does not depend on the pointing direction of the beam. Hence, the received power of the $i$-th VCSEL is maximized when the beam axis passes through the PD, and, thus, the pointing direction $\mathbf{b}_i$ of the $i$-th VCSEL at this instant, which is known from its sweep, satisfies}
\begin{small}
\begin{equation}
\mathbf{b}_i=\frac{\mathbf{r}-\mathbf{s}_i}{\|\mathbf{r}-\mathbf{s}_i\|},
\label{eq:pointing_direction}
\end{equation}    
\end{small}
{which implies that}
\begin{small}
\begin{equation}
\mathbf{Q}_i(\mathbf{r}-\mathbf{s}_i)=\mathbf{0},
\label{eq:pointing_constraint}
\end{equation}    
\end{small}
{where $\mathbf{Q}_i=\mathbf{I}_3-\mathbf{b}_i\mathbf{b}_i^T$ denotes the orthogonal projector onto the plane perpendicular to $\mathbf{b}_i$, and, thus, \eqref{eq:pointing_constraint} provides two independent linear equations in $\mathbf{r}$ for each VCSEL. Therefore, by collecting the three VCSELs, it follows that}
\begin{small}
\begin{equation}
\left(\sum_{i=1}^{3}\mathbf{Q}_i\right)\mathbf{r}=\sum_{i=1}^{3}\mathbf{Q}_i\mathbf{s}_i,
\label{eq:position_system}
\end{equation}    
\end{small}
{where $\sum_{i=1}^{3}\mathbf{Q}_i$ is nonsingular, since $\mathbf{v}^T\big(\sum_{i=1}^{3}\mathbf{Q}_i\big)\mathbf{v}=\sum_{i=1}^{3}\|\mathbf{Q}_i\mathbf{v}\|^2$ vanishes only for a vector $\mathbf{v}$ that is parallel to all the three pointing directions, which cannot all be parallel under the first condition, since, otherwise, the three VCSELs would lie on the same line through the UE. As a result, the UE position is uniquely obtained in closed form as}
\begin{small}
\begin{equation}
\hat{\mathbf{r}}=\left(\sum_{i=1}^{3}\mathbf{Q}_i\right)^{-1}\sum_{i=1}^{3}\mathbf{Q}_i\mathbf{s}_i,
\label{eq:position_closed_form}
\end{equation}    
\end{small}
{without any mirror-symmetric candidate.}
 
With the UE position $\hat{\mathbf{r}}$ determined, the final step is to recover the orientation vector. From \eqref{modelp}, it follows that the orientation vector appears only through its inner product with the direction from the UE to each VCSEL. Hence, for each VCSEL, we define
\begin{small}
\begin{equation}
\hat{\mathbf{u}}_i=\frac{\mathbf{s}_i-\hat{\mathbf{r}}}{\|\mathbf{s}_i-\hat{\mathbf{r}}\|},
\end{equation}   
\end{small}
which represents the unit vector from the UE to the $i$-th VCSEL. {Since the distances $d_i=\|\mathbf{s}_i-\hat{\mathbf{r}}\|$ are thus determined, by setting $c_i=P_i/\beta_i(d_i)$ and invoking \eqref{modelp}}, the orientation vector will satisfy the linear system
\begin{small}
\begin{equation}
\underbrace{\begin{bmatrix}
\hat{\mathbf{u}}_1^{T}\\ \vdots\\ \hat{\mathbf{u}}_{N_{\mathrm{V}}}^{T}
\end{bmatrix}}_{{\mathbf{U}}\in\mathbb{R}^{N_{\mathrm{V}}\times 3}}\mathbf{n}
=
\underbrace{\begin{bmatrix}
{c_1}\\ \vdots\\ {c_{N_{\mathrm{V}}}}
\end{bmatrix}}_{\mathbf{c}},
\label{eq:solve_n}
\end{equation}
\end{small}which establishes a direct relation between the orientation vector and the normalized measurements once the position has been fixed. Thus, considering that $N_{\mathrm{V}}=3$, the system \eqref{eq:solve_n} reduces to a $3\times 3$ linear system, and, if $\det{\mathbf{U}}\neq 0$, the orientation vector of the user can be written in closed form as
\begin{small}
\begin{equation}
\tilde{\mathbf{n}}={\mathbf{U}}^{-1}\mathbf{c},
\end{equation}    
\end{small}which concludes the proof.
\end{IEEEproof}
\begin{remark}
{The condition $\det\mathbf{U}\neq0$ holds if and only if the UE does not lie on the plane defined by the three VCSELs, since $\det[\mathbf{s}_1-\mathbf{r},\mathbf{s}_2-\mathbf{r},\mathbf{s}_3-\mathbf{r}]$ vanishes precisely when these points are coplanar. Hence, when the three VCSELs are integrated at non-collinear positions on the same LeRIS panel, this plane coincides with the wall on which the panel is mounted, and, thus, $\det\mathbf{U}\neq0$ holds for every UE position inside the room, which enables a unique recovery of both position and orientation.}
\end{remark}
\begin{remark}
{Since VCSELs support transverse modes with different divergences, depending on the aperture geometry and the bias conditions, the wider mode enables the fast acquisition of the UE during the sweep, while the narrower mode sharpens the received-power peak and, thus, improves the accuracy of the pointing directions $\mathbf{b}_i$, consistently with the impact of the beam divergence on the performance of VCSEL-based systems analyzed in \cite{Zeng2021_VCSEL_BeamActivation}. In practice, finite angular sweep resolution and noise in the received-power peak detection perturb the pointing directions. Specifically, by defining $\mathbf A=\sum_{i=1}^{3}\mathbf Q_i$ and considering small perturbations $\Delta\mathbf b_i$ orthogonal to $\mathbf b_i$, first-order linearization of \eqref{eq:position_system} yields
\begin{small}
\begin{equation}
\Delta\mathbf r\approx\mathbf A^{-1}\sum_{i=1}^{3}d_i\Delta\mathbf b_i,
\label{eq:direction_perturbation}
\end{equation}    
\end{small}
and, hence,
\begin{small}
\begin{equation}
\|\Delta\mathbf r\|\lesssim\frac{\sum_{i=1}^{3}d_i\|\Delta\mathbf b_i\|}{\lambda_{\min}(\mathbf A)}.
\label{eq:direction_error_bound}
\end{equation}    
\end{small}
The position error therefore depends on the angular accuracy and the conditioning of $\mathbf A$, while the subsequent orientation recovery is affected by RSS noise and the conditioning of $\mathbf U$. In particular, for zero-mean independent components of each $\Delta\mathbf b_i$ with variance $\sigma_b^2$, e.g., $\sigma_b^2=\Delta_\theta^2/12$ for a sweep with angular step $\Delta_\theta$, \eqref{eq:direction_perturbation} yields the position RMSE $\sigma_b\big(\operatorname{tr}\big(\mathbf A^{-1}\big(\sum_{i=1}^{3}d_i^2\mathbf Q_i\big)\mathbf A^{-1}\big)\big)^{1/2}$, which does not depend on the PD orientation, since neither $\mathbf A$ nor $d_i$ depends on $\mathbf n$. Moreover, for the parameters of Table~\ref{table1}, a deviation of $1^\circ$ from the UE direction reduces the received power by approximately 19\%, whereas the RSS noise remains below 0.1\% of the received power inside the room, and, thus, the noise can only alter the detected peak between two pointing directions that are almost equally close to the UE direction, which keeps the angular error within the sweep step.}
\label{rem:pointing}
\end{remark}
\begin{remark}
Compared to LED-based LeRIS \cite{R2}, which requires at least four optical sources to resolve a receiver's 3D position, the proposed VCSEL-based framework achieves unique localization with only three, owing to the high directionality of VCSEL beams{, which provides the pointing direction of each VCSEL in addition to its RSS value}, highlighting the efficiency of the proposed LeRIS architecture.
\end{remark}
Proposition~\ref{prop1} establishes that the UE position and orientation can be uniquely recovered from {three VCSELs} in the absence of noise. In practice, however, receiver noise introduces uncertainty into the RSS measurements and consequently limits the accuracy of the jointly estimated UE state. To quantify this fundamental limitation, we derive the position error bound (PEB) of the proposed localization framework by exploiting the information provided by all visible VCSEL observations while accounting for the unknown PD orientation.
 
\begin{proposition}
\label{prop:joint_fim_peb}
Let
{
\begin{small}
\begin{equation}
\boldsymbol{\xi}_k=[\mathbf{r}_k^T,\varphi_k,\vartheta_k]^T,
\label{eq:joint_state_vector}
\end{equation}    
\end{small}
}denote the unknown state of UE $k$, where
$\mathbf r_k\in\mathbb R^3$ is its position, while
$\varphi_k$ and $\vartheta_k$ denote the azimuth and elevation
angles of the PD orientation, respectively. The corresponding unit-normal vector of the PD is expressed as
\begin{small}
\begin{equation}
\mathbf n_k
=
\begin{bmatrix}
\cos\vartheta_k\cos\varphi_k\\
\cos\vartheta_k\sin\varphi_k\\
\sin\vartheta_k
\end{bmatrix},
\label{eq:pd_orientation_vector}
\end{equation}    
\end{small}
where the elevation angle $\vartheta_k$ is measured from the
horizontal plane. {Let $\mathcal V_k^{(L)}$ denote the set of all VCSELs visible to UE $k$ across the $L$ participating LeRIS panels, with $N_{\mathrm{V}}=|\mathcal V_k^{(L)}|$.} Under the
RSS model in \eqref{modelp}, the mean observation associated
with VCSEL $i\in\mathcal V_k^{(L)}$ is
\begin{small}
\begin{equation}
\mu_{i,k}(\boldsymbol{\xi}_k)
=
\beta_i(d_{i,k})
\left(
\mathbf n_k^{T}\mathbf u_{i,k}
\right).
\label{eq:fim_mean_rss}
\end{equation}    
\end{small}
The corresponding RSS measurement is modeled as
\begin{small}
\begin{equation}
P_{r,i,k}
=
\mu_{i,k}(\boldsymbol{\xi}_k)+\nu_{i,k},
\qquad
\nu_{i,k}
\sim
\mathcal N\left(0,P_{n,i,k}^{2}\right),
\label{eq:fim_rss_observation}
\end{equation}    
\end{small}
where $P_{n,i,k}$ is the standard deviation of the equivalent
optical noise defined in
\eqref{eq:equivalent_optical_noise_variance}.
 
Assuming statistically independent VCSEL observations, the Fisher information matrix associated with the complete UE state is
\begin{small}
 \begin{equation}
\mathbf J_k^{(L)}
=
\sum_{i\in\mathcal V_k^{(L)}}
{\frac{1+\varepsilon_{i,k}}{P_{n,i,k}^{2}}}
\mathbf g_{i,k}\mathbf g_{i,k}^{T},
\label{eq:joint_state_fim}
\end{equation}   
\end{small}
where
\begin{small}
\begin{equation}
\mathbf g_{i,k}
\triangleq
\nabla_{\boldsymbol{\xi}_k}
\mu_{i,k}(\boldsymbol{\xi}_k){,}
\label{eq:fim_measurement_gradient}
\end{equation}    
\end{small}

{while $\varepsilon_{i,k}=\varsigma_{i,k}^{2}/(2P_{n,i,k}^{2})$, with $\varsigma_{i,k}=2B_o\left(q/R_{\mathrm{PD}}+\mathrm{RIN}\,\mu_{i,k}\right)$, accounts for the dependence of the noise variance on the UE state.}
Consequently, the three-dimensional position error bound of
UE $k$ is
\begin{small}
  \begin{equation}
\mathrm{PEB}_k^{(L)}
=
\sqrt{
\operatorname{tr}
\left(
\left[
\left(
\mathbf J_k^{(L)}
\right)^{-1}
\right]_{1:3,1:3}
\right)
}.
\label{eq:position_error_bound}
\end{equation}  
\end{small}

\end{proposition}
 
\begin{IEEEproof}
The proof is provided in Appendix~\ref{app:joint_fim_proof}.
\end{IEEEproof}
\begin{remark}
{Since \eqref{eq:joint_state_fim} can be written as $\mathbf J_k^{(L)}=\mathbf H_k^T\mathbf W_k\mathbf H_k$, where $\mathbf H_k\in\mathbb R^{N_{\mathrm{V}}\times5}$ is the matrix whose rows are $\mathbf g_{i,k}^T$, $i\in\mathcal V_k^{(L)}$, and $\mathbf W_k$ is the diagonal matrix of the corresponding positive weights, $\mathbf J_k^{(L)}$ is nonsingular, and, thus, the PEB in \eqref{eq:position_error_bound} is finite, if and only if $\mathbf H_k$ has full column rank, i.e., if and only if the visible VCSELs contribute independent RSS values, which coincides with the condition of Lemma~\ref{lemma1} and guarantees the local identifiability of the UE state \cite{Rothenberg1971}. Similarly, for correlated measurement errors with a constant positive-definite covariance matrix $\boldsymbol{\Sigma}_k$, \eqref{eq:joint_state_fim} generalizes to $\mathbf J_k^{(L)}=\mathbf H_k^T\boldsymbol{\Sigma}_k^{-1}\mathbf H_k$, which is also nonsingular if and only if $\mathbf H_k$ has full column rank. Moreover, when all the visible VCSELs belong to the same LeRIS panel, the mirror image of the UE with respect to the mounting wall, together with the correspondingly mirrored PD orientation, yields the same RSS values, and, thus, it is excluded, since it lies behind the wall. Finally, since}
\begin{equation}
\begin{aligned}
\varepsilon_{i,k}&=\frac{2B_o\left(q+\mathrm{RIN}\,R_{\mathrm{PD}}\mu_{i,k}\right)^2}{A_K+2qR_{\mathrm{PD}}\mu_{i,k}+\mathrm{RIN}\,R_{\mathrm{PD}}^2\mu_{i,k}^2}\\
&\leq\bar{\varepsilon}=\max\left\{\frac{2q^2B_o}{A_K},\,2B_o\mathrm{RIN}\right\}
\end{aligned}
\end{equation}
{for every UE state, the dependence of the noise variance on the UE state scales the contribution of each VCSEL observation to $\mathbf J_k^{(L)}$ by at most $1+\bar{\varepsilon}$, which equals $1+6.3\times10^{-13}$ for the parameters of Table~\ref{table1}, and, thus, the PEB in \eqref{eq:position_error_bound} differs by less than $3.2\times10^{-13}$ in relative terms from the one obtained for $\varepsilon_{i,k}=0$.}
\end{remark}
 
\subsection{VCSEL-based Mapping}
 
\begin{figure}
    \centering
    \includegraphics[width=0.9\columnwidth]{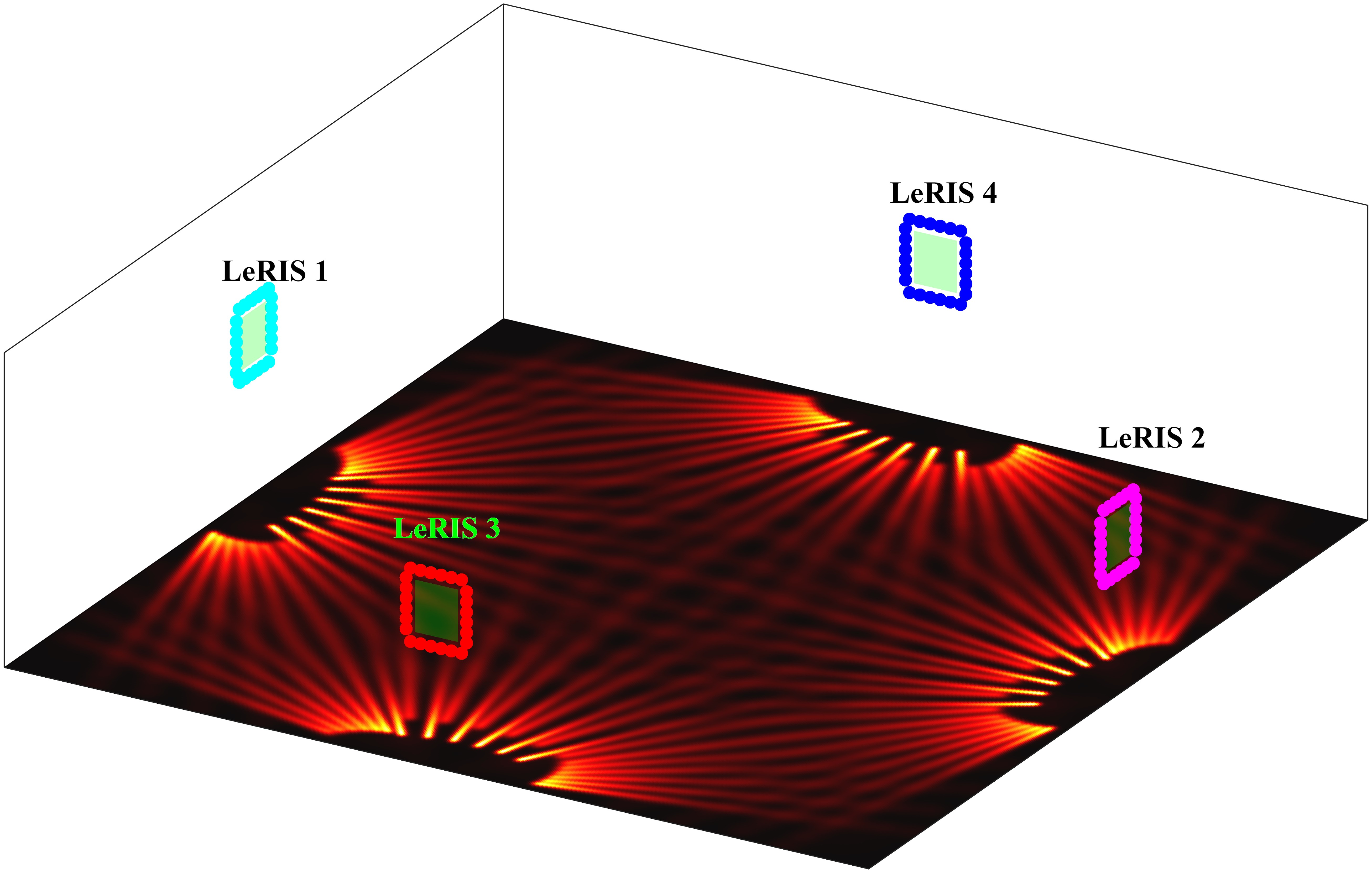}
    \caption{VCSEL coverage from multiple LeRISs}
    \label{fig:heatmap}
\end{figure}
 
After the UE reports its position, the AP configures the PWE by selecting a LeRIS route toward the UE to provide high data rate connectivity in the mmWave band. The effectiveness of the selected LeRIS route, however, depends not only on accurate localization, but also on the guarantee that each selected LeRIS-UE link is unobstructed. In more detail, in the mmWave band, even small obstacles can block the LoS and significantly degrade the end-to-end rate, which makes it necessary to complement localization with a mechanism that reveals the presence of blocking objects. In this direction, VCSEL-based LeRISs can also act as sensing anchors, since the high directionality of VCSELs ensures that, whenever an object lies along the beam axis, a distinct portion of the signal is reflected back toward the LeRIS due to its narrow angular footprint. In more detail, as the VCSELs in the proposed LeRIS architecture rotate over small angular ranges to provide coverage, the resulting reflections can be captured by the co-located PDs and associated with the corresponding angular position of the VCSEL, thereby enabling the surrounding space to be mapped in a structured manner that reveals potential obstructions. Finally, as shown in Fig.~\ref{fig:heatmap}, when multiple LeRIS panels illuminate the environment, their rotating VCSEL beams span the space in a complementary manner, which allows the surrounding region to be simultaneously mapped and monitored for obstacles, while also supporting localization. In this way, the same VCSEL-based infrastructure that enables closed-form localization also extends the functionality of PWEs by providing sensing capabilities, ensuring that the configured routes are unobstructed.
 
To determine if a transmitting node at position $(x_T, y_T, z_T)$ and a receiving node at $(x_R, y_R, z_R)$ are obstructed through the proposed sensing scheme, we first note that the LoS segment connecting them can be written as
\begin{small}
\begin{equation}
\mathbf{r}(t)=\mathbf{P}_0+t\,\mathbf{d}_s,
\quad 0\leq t\leq 1,
\label{eq:los_segment}
\end{equation}    
\end{small}where $\mathbf{P}_0=(x_T,y_T,z_T)$, and ${\mathbf{d}_s}=(x_R-x_T, y_R-y_T, z_R-z_T)$ is the direction vector of the segment. During the VCSEL angular sweep, beams with unit direction $\mathbf{d}=[\cos\theta\cos\phi,\ \cos\theta\sin\phi,\ \sin\theta]$ generate reflection points that are registered as $\mathbf{r}=\mathbf{P}+d_V\cdot\mathbf{d}$ from the beam origin $\mathbf{P}$ at range $d_V$, where the corresponding round-trip time delay is given by $\tau = 2d_V/c$, where $c$ is the speed of light.
 
{To capture the mapping-based link-feasibility decision in the communication model, a binary indicator $\chi_s\in\{0,1\}$ is introduced for each segment $s$, where $\chi_s=0$ denotes a segment that is identified as obstructed from the registered reflection points, and $\chi_s=1$ denotes a segment for which no obstruction is identified.} Therefore, in the following proposition, we establish the condition under which this LoS segment is considered obstructed.
\begin{proposition}
\label{prop:blockage}
{Let the registered reflection points of each detected obstacle describe its surface through planar faces, where each face $F$ is the convex hull $\mathcal{H}_F$ of at least three non-collinear reflection points that lie on the same plane, and let $\mathbf{n}_F$ denote the unit normal of this plane, which contains the reflection point $\mathbf{r}_F\in\mathcal{H}_F$. Moreover, let $s_F(\mathbf{x})=\mathbf{n}_F^T(\mathbf{x}-\mathbf{r}_F)$ denote the signed distance of a point $\mathbf{x}$ from the plane of $F$. If, for at least one face $F$, it holds that $\mathbf{n}_F^T\mathbf{d}_s\neq0$, $s_F(\mathbf{P}_0)\,s_F(\mathbf{P}_0+\mathbf{d}_s)\leq0$, and $\mathbf{P}_0+t_F\mathbf{d}_s\in\mathcal{H}_F$, where $t_F=\mathbf{n}_F^T(\mathbf{r}_F-\mathbf{P}_0)/(\mathbf{n}_F^T\mathbf{d}_s)$, then the LoS segment in \eqref{eq:los_segment} is obstructed, and $\chi_s=0$, while, otherwise, the segment is identified as unobstructed, and $\chi_s=1$.}
\end{proposition}
\begin{IEEEproof}
{Since $s_F(\mathbf{P}_0+t\,\mathbf{d}_s)=s_F(\mathbf{P}_0)+t\,\mathbf{n}_F^T\mathbf{d}_s$ is affine in $t$, with $\mathbf{n}_F^T\mathbf{d}_s\neq0$, the condition $s_F(\mathbf{P}_0)\,s_F(\mathbf{P}_0+\mathbf{d}_s)\leq0$ holds if and only if the segment crosses the plane of $F$ at $t_F\in[0,1]$, i.e., at the point $\mathbf{P}_0+t_F\mathbf{d}_s$, while $\mathbf{P}_0+t_F\mathbf{d}_s\in\mathcal{H}_F$ holds if and only if this point lies on the face $F$. Therefore, the two conditions establish that the finite LoS segment itself crosses the surface of the detected obstacle, which concludes the proof.}
\end{IEEEproof}
\begin{remark}
{The criterion of Proposition~\ref{prop:blockage} relies on the reflection points being spatially resolved, which is enabled by the high directionality of the VCSELs, since each reflection is associated with a known narrow beam direction and a ToA-based range, and, thus, the reflection points registered during the sweep directly sample the faces of the obstacles. It should be noted that, since the two nodes lie outside the obstacles, a segment that passes through an obstacle enters its interior by crossing at least one of its faces with $\mathbf{n}_F^T\mathbf{d}_s\neq0$, and, hence, when the faces of the obstacles are fully mapped, the condition of Proposition~\ref{prop:blockage} is also necessary, and, thus, $\chi_s$ coincides with the actual availability of the segment. On the contrary, missed or spurious reflections may result in erroneous decisions, i.e., an obstructed segment that is identified as unobstructed or vice versa, whose effect is evaluated in Section~\ref{sec:IV}.}
\end{remark}
 
\subsection{LeRIS-based Communication for Multiple UEs}
{After localization and mapping, the AP selects a route toward each UE using the link states supplied by the VCSEL-based sensing procedure. Specifically, let $\chi_s$ be the sensing-based indicator of Proposition~\ref{prop:blockage}, and let $\chi_s^{\rm act}$ denote the actual physical availability of segment $s$. The two indicators coincide in the ideal sensing case but may differ when blockage information is imperfect. For a candidate route $\pi$, its sensing-based feasibility is given by $\chi_{\pi,k}=\prod_{s\in S(\pi)}\chi_s$, with $S(\pi)$ denoting its set of segments. Consequently, the set of routes admitted by the sensing information is $\mathcal P_k^{\rm sens}=\{\pi:\chi_{\pi,k}=1\}$. Within this set, the nominal path losses and nominal beamforming gains are used to rank the candidate routes. Thus, by defining the nominal route rate as}
\begin{small}
\begin{equation}
R_{\pi,k}^{\rm nom}
=B\log_2\!\left(1+l_p^{\rm nom}(\pi)G_tG_r
\gamma_tG_{\pi,k}^{\rm nom}\right),
\label{eq:nominal_route_spectral_efficiency}
\end{equation}    
\end{small}{the selected route is}
\begin{small}
\begin{equation}
\pi_k^\star
=\underset{\pi\in\mathcal P_k^{\rm sens}}{\arg\max}
R_{\pi,k}^{\rm nom}.
\label{eq:max}
\end{equation}    
\end{small}{Here, $l_p^{\rm nom}(\pi)$ and $G_{\pi,k}^{\rm nom}$ are calculated using the nominal panel positions and ideal phase profiles with unit reflection efficiency, without the hardware errors or panel displacements. Accordingly, neither the realized phase perturbations nor the offsets influence the route-selection decision. Once the route has been selected, its achieved rate is evaluated as}
\begin{small}
\begin{equation}
R_k
=\chi_{\pi_k^\star,k}^{\rm act}
B\log_2\!\left(1+l_p^{\rm act}(\pi_k^\star)
G_tG_r\gamma_tG_{\pi_k^\star,k}^{\rm act}\right),
\label{eq:route_spectral_efficiency}
\end{equation}    
\end{small}{where $\chi_{\pi,k}^{\rm act}=\prod_{s\in S(\pi)}\chi_s^{\rm act}$, while $l_p^{\rm act}$ and $G_{\pi,k}^{\rm act}$ account for the actual propagation geometry and the applied phases. If $\mathcal P_k^{\rm sens}$ is empty, $R_k$ is set to zero. This distinction accounts for routes that are mistakenly classified as feasible or unavailable under imperfect sensing, without allowing hardware errors or offsets to influence the sensing-based route decision.}

When multiple UEs are simultaneously present in the PWE, their end-to-end performance depends not only on the selected LeRIS routes but also on how the available transmission time is shared among them. Thus, since {the achieved rate $R_k$ is obtained from \eqref{eq:route_spectral_efficiency} after the sensing-based route selection in \eqref{eq:max}}, to enable fair access among the $K$ UEs of the set $\mathcal{K}=\{1,\ldots,K\}$ within a TDMA framework, an optimization problem is formulated to maximize the minimum user rate, i.e.,
\begin{small}
 \begin{equation}
\begin{aligned}
\max_{\{\tau_k\}}\quad & R_{\min}\\
\text{s.t.}\quad & \tau_kR_k\geq R_{\min},\quad\forall k\in\mathcal{K},\\
& \textstyle\sum_{k=1}^{K}\tau_k=1,\\
& 0\leq\tau_k\leq1,\quad\forall k\in\mathcal{K},
\end{aligned}
\label{eq:TDMA}
\end{equation}   
\end{small}where $\tau_k$ denotes the time allocation of the $k$-th UE, and, thus, $\tau_kR_k$ denotes its TDMA-adjusted user rate. As it can be observed, the problem in \eqref{eq:TDMA} is convex, since the objective is linear and the constraints are affine in the optimization variables, which guarantees that a globally optimal solution can be efficiently obtained \cite{bozanis_opt}. {In particular, when $R_k>0$ for all $k$, all the rate constraints hold with equality at the optimum, and, thus, the optimal time allocation is given in closed form by $\tau_k=R_{\min}/R_k$, where $R_{\min}=\big(\sum_{j=1}^{K}1/R_j\big)^{-1}$. Moreover, the configuration of the LeRIS panels follows directly from the spatial information acquired in the optical domain, which complements joint RIS and resource-optimization frameworks that configure the phase shifts through the knowledge of the cascaded channels and iterative optimization \cite{Xiu2021RISNOMA}.} It should be noted that the {rate} $R_k$ in \eqref{eq:TDMA} captures the {realized performance of the sensing-selected route, including actual blockage, cascaded path loss, and beamforming imperfections, as established in \eqref{eq:max} and \eqref{eq:route_spectral_efficiency}. The route decision itself does not depend on hardware phase errors or panel offsets.} Therefore, the obtained time allocation directly reflects the localization, mapping, and routing outcomes of the proposed framework.

\section{Simulation Results}\label{sec:IV}

In this section, we evaluate the performance of a multi-user indoor system with four LeRIS panels, whose corresponding parameter values for the VCSELs are listed in Table~\ref{table1}, while the communication parameters are provided in Table~\ref{table2}. In more detail, each LeRIS is equipped with 24 VCSELs placed along its perimeter{, i.e., along a square frame of side 0.7~m centered at the LeRIS}, with each VCSEL independently rotating its emitted beam to scan a common azimuthal sector of $120^\circ$ and an elevation span of $60^\circ$, thereby providing wide angular coverage and allowing multiple VCSEL beams to illuminate the same UE. As a result, the deployed beams complement one another to span the served area, enabling user localization and obstacle sensing, while simultaneously supporting LeRIS-based communication. {Moreover, the UE coordinates are uniformly sampled within $x_u\in[0,10]$~m, $y_u\in[0,10]$~m, and $z_u\in[0.5,1.5]$~m, excluding the volumes occupied by the obstacles, where the range of $z_u$ covers hand-held devices of both sitting and standing users, while the azimuth and elevation orientation angles $\varphi_k$ and $\vartheta_k$ of the UEs are uniformly distributed over $[0,2\pi)$ and $[0,\pi/2]$, respectively, which covers all the upward-facing orientations of the device and, thus, the entire range of the measurement-based orientation model in \cite{DehghaniSoltani2019}, where the azimuth angle is also uniformly distributed.} In addition to the UEs, $O_b$ cuboid objects of dimensions $1 \times 1 \times 1.6$~m are uniformly distributed within the PWE, so that the generated layouts capture random blockage events consistent with the user height. {Regarding the communication evaluation, an ideal and a realistic case are compared using the same UE positions and obstacle realizations, where the ideal case assumes unit reflection-power efficiency, nominal panel locations, exact phase actuation, and correct sensing-based link states. On the contrary, in the realistic case, the reflection-power factor is set to $\eta_{\rm eff}=|\Gamma|^2=0.81$, corresponding to a reference unit-cell reflection amplitude $|\Gamma|=0.9$, independent von Mises phase errors with $\kappa_{\rm hw}=10$ are imposed on every participating LeRIS \cite{Wang2021VonMises}, and the position of each panel is displaced by $\Delta_o=0.01$~m parallel to its mounting wall, where the RIS configuration is derived from the nominal geometry, while the actual gain and path loss are evaluated using the displaced geometry. Furthermore, imperfect sensing is represented by potentially erroneous segment-availability decisions, i.e., false-feasible and false-blocked decisions, which are characterized by $p_{\rm FP}=\Pr(\chi_s=1\mid\chi_s^{\rm act}=0)=0.10$ and $p_{\rm FN}=\Pr(\chi_s=0\mid\chi_s^{\rm act}=1)=0.10$, respectively.} Finally, to quantify the system performance under realistic conditions, we conduct a Monte Carlo simulation with $N_{\mathrm{MC}}=10^5$ iterations, where, in each iteration, both the UE positions and orientations, as well as the locations of the blocking objects, are randomly generated{, while the UE position and orientation are estimated through Lemma~\ref{lemma1}, or through Proposition~\ref{prop1} whenever Lemma~\ref{lemma1} cannot be applied or its RMSE in Remark~\ref{rem:pointing} is lower than the PEB in \eqref{eq:position_error_bound}, both evaluated at the estimated UE state, and employed for the steering of the final LeRIS of the selected route}.
 
\begin{table}[]
	\renewcommand{\arraystretch}{1.25}
	\caption{\textsc{Simulation Parameters for VCSELs}}
	\label{PWE_char2}
	\centering
	\begin{tabular}{ll}
		\hline
		\bfseries Parameter & \bfseries Value \\
		\hline\hline
        LeRIS 1 {Center} coordinates & $\left(0, 5, 1.5\right)$ \\
		LeRIS 2 {Center} coordinates & $\left(10, 5, 1.5\right)$ \\
  	    LeRIS 3 {Center} coordinates & $\left(5, 0, 1.5\right)$ \\
        LeRIS 4 {Center} coordinates & $\left(5, 10, 1.5\right)$ \\
        PD area & $A_\mathrm{PD} = 1$ cm${}^2$ \\
        Transmit power & $P_{t,i} = 10$ mW \\
        Beam waist & ${w_0} = 5.6$ $\mu$m \\
        Speed of light & $c = 3\times10^8 \mathrm{m/s}$ \\
        VCSEL wavelength & $\lambda_o = 950$ nm \\
        {Tone-filter noise bandwidth} & {$B_o = 1$ kHz} \\
        Relative intensity noise & $\mathrm{RIN} = -155$ {dB/Hz} \\
        Preamplifier noise figure & $F_{n} = 5$ dB \\
        PD FoV & $\Psi = 90\degree$ \\
Electrical noise variance
&
{$\sigma_{I,i,k}^{2}=B_oS_{i,k}$, $S_{i,k}$ in \eqref{eq:electrical_noise_psd}} \\
        Load resistance & $R_\mathrm{L} = 50\Omega$ \\
        Absolute temperature & $T = 300 \mathrm{K}$ \\
        Boltzmann constant & $k_B = 1.380649 \times10^{-23}\,{\mathrm{J/K}}$ \\
        PD responsivity & $R_{\mathrm{PD}} = 0.7$ A/W \\
		\hline
	\end{tabular}
    \label{table1}
\end{table}
 
\begin{table}[]
	\renewcommand{\arraystretch}{1.25}
	\caption{\textsc{Simulation Parameters for mmWave communication}}
	\label{PWE_char}
	\centering
	\begin{tabular}{ll}
		\hline
		\bfseries Parameter & \bfseries Value \\
		\hline\hline
        {mmWave wavelength} & $ \lambda_m = 10^{-2} $ m \\
        {AP antenna gain} & $G_t = 10$ dB \\
        {UE antenna gain} & $G_r = 0$ dB \\
        Reference distance & $d_{r,0} = 1$ m \\
        Path loss exponent & ${n_s} = 2$ \\
        {Element side length} & $D = \frac{\lambda_m}{2}$ \\
        {Reflection-power factor (ideal/realistic)} & {$\eta_{\mathrm{eff}}=1/0.81$} \\
        {Phase-error concentration (realistic)} & {$\kappa_{\rm hw}=10$} \\
        {Panel displacement (realistic)} & {$\Delta_o=0.01~\mathrm{m}$} \\
        {Sensing-error parameters (realistic)} & {$p_{\rm FP}=p_{\rm FN}=0.10$} \\
        {System bandwidth} & {$B=100~\mathrm{MHz}$} \\
        {UE receiver noise figure} & {$\mathrm{NF}=5$~dB} \\
        AWGN variance & {$\sigma^2_{k}=-119$~dBW} \\
		\hline
	\end{tabular}
    \label{table2}
\end{table}

\begin{figure}[t]
  \centering
  \begin{subfigure}[t]{0.95\linewidth}
    \centering
    \begin{tikzpicture}[scale=0.52]
    \foreach \g in {2,4,6,8} {\draw[gray!25] (\g,0)--(\g,10); \draw[gray!25] (0,\g)--(10,\g);}
    \draw[line width=1pt] (0,0) rectangle (10,10);
    \filldraw[red!80!black, fill=white, line width=1pt] (0,5) circle (0.22);
    \filldraw[magenta, fill=white, line width=1pt] (10,5) circle (0.22);
    \filldraw[green!50!black, fill=white, line width=1pt] (5,0) circle (0.22);
    \filldraw[blue, fill=white, line width=1pt] (5,10) circle (0.22);
    \node[red!80!black, right, font=\scriptsize\bfseries] at (0.3,5) {LeRIS 1};
    \node[magenta, left, font=\scriptsize\bfseries] at (9.7,5) {LeRIS 2};
    \node[green!50!black, above, font=\scriptsize\bfseries] at (5,0.3) {LeRIS 3};
    \node[blue, below, font=\scriptsize\bfseries] at (5,9.7) {LeRIS 4};
    \draw[line width=1.2pt] (4.3,5)--(5.7,5); \draw[line width=1.2pt] (5,4.3)--(5,5.7);
    \node[font=\tiny] at (6.3,5) {$0^\circ$};
    \node[font=\tiny] at (5,6.05) {$90^\circ$};
    \node[font=\tiny] at (3.6,5) {$180^\circ$};
    \node[font=\tiny] at (5,3.95) {$270^\circ$};
    \node[blue, font=\scriptsize\bfseries] at (5.75,4.45) {UE};
    \end{tikzpicture}
    \caption{}
    \label{fig:user_orient}
  \end{subfigure}
  \vspace{2mm}
  \begin{subfigure}[t]{0.95\linewidth}
    \centering
    \begin{tikzpicture}
    \begin{axis}[
        width=0.95\linewidth,
        xlabel = {$\varphi_k$},
        ylabel = {$\mathrm{RMSE}$(cm)},
        xmin = 0, xmax = 360,
        ymode = log,
        ymin = 0.1, ymax = 1000,
        xtick = {0,60,120,180,240,300,360},
        xticklabels = {$0^\circ$, $60^\circ$, $120^\circ$, $180^\circ$,
                       $240^\circ$, $300^\circ$, $360^\circ$},
        ytick = {0.1,1,10,100,1000},
        grid = major,
        unbounded coords=jump,
        filter discard warning=false,
        legend cell align = {left},
        legend columns=2,
        transpose legend=false,
        legend style = {
            font=\scriptsize,
            at={(0.5,0.97)},
            anchor=north,
            draw=black,
            fill=white,
            cells={anchor=west}
        }
    ]
    \addplot[blue, line width=1pt]
    table[x index=0, y index=1] {Fig3b_L1_RMSE.dat};
    \addlegendentry{$L=1$}
    \addplot[black, line width=1pt]
    table[x index=0, y index=1] {Fig3b_L2_RMSE.dat};
    \addlegendentry{$L=2$}
    \addplot[red, line width=1pt]
    table[x index=0, y index=1] {Fig3b_L4_RMSE.dat};
    \addlegendentry{$L=4$}
    \addplot[blue, only marks, mark=*, mark size=1.5pt, mark repeat=9, mark phase=1]
    table[x index=0, y index=1] {Fig3b_L1_PEB.dat};
    \addlegendentry{$L=1$ PEB/Rem.~3}
    \addplot[black, only marks, mark=*, mark size=1.5pt, mark repeat=9, mark phase=4]
    table[x index=0, y index=1] {Fig3b_L2_PEB.dat};
    \addlegendentry{$L=2$ PEB/Rem.~3}
    \addplot[red, only marks, mark=*, mark size=1.5pt, mark repeat=9, mark phase=7]
    table[x index=0, y index=1] {Fig3b_L4_PEB.dat};
    \addlegendentry{$L=4$ PEB/Rem.~3}
    \end{axis}
    \end{tikzpicture}
    \caption{}
    \label{fig:phi_vs_error}
  \end{subfigure}
  \caption{(a) Considered LeRIS deployment and UE-orientation reference, (b) Position RMSE {(lines) and its theoretical value (markers), i.e., the PEB in \eqref{eq:position_error_bound} or the RMSE in Remark~\ref{rem:pointing},} versus {$\varphi_k$} for different numbers of participating LeRIS panels $L$.}
  \label{fig:orientation_and_error}
\end{figure}
 
Fig.~\ref{fig:user_orient} illustrates the considered UE-orientation geometry, while Fig.~\ref{fig:phi_vs_error} presents the position root-mean-squared error (RMSE) and {its theoretical value, i.e., the PEB in \eqref{eq:position_error_bound} or the RMSE in Remark~\ref{rem:pointing} of the employed method,} as a function of {$\varphi_k$} for different numbers of participating LeRIS panels{, where the UE is located at $(5,5,1.5)$~m, i.e., at equal distances from the four LeRIS panels, and the PD elevation is set to $\vartheta_k=0$, in order to isolate the dependence of the localization accuracy on the azimuth orientation of the UE, while all the five parameters of $\boldsymbol{\xi}_k$ are estimated}. {In addition, the RSS measurements are generated, and the PEB is computed, with the equivalent noise variance in \eqref{eq:equivalent_optical_noise_variance} and \eqref{eq:electrical_noise_psd}, evaluated at the true UE state.} {Moreover, the LM iteration of Appendix~\ref{App1} starts from a random point, with a standard deviation of 0.15~m per coordinate around the true UE position, i.e., its previous estimate, and is restarted from points located 0.1~m and 0.2~m away whenever $J(\hat{\mathbf{r}})$ in \eqref{minproblem} exceeds ten times $\frac{1}{2}\sum_{i}P_{n,i,k}^{2}$, which indicates a local minimum. Accordingly, the RSS-based estimator of Lemma~\ref{lemma1} is employed, while Proposition~\ref{prop1}, with a sweep of angular step $\Delta_\theta=1^\circ$, takes over whenever Lemma~\ref{lemma1} cannot be applied or its RMSE in Remark~\ref{rem:pointing} is lower than the PEB in \eqref{eq:position_error_bound}. As it can be observed, for $L=1$, the PEB of the RSS-based estimator exceeds 1~m when the PD faces the panel, since a lateral or vertical displacement of the UE and a tilt of its PD then yield almost identical RSS variations, whereas Proposition~\ref{prop1}, whose pointing directions do not depend on the PD orientation, takes over for $|\varphi_k|\leq11^\circ$ and limits the position RMSE to 16.8~cm. As $\varphi_k$ increases, the different incidence angles of the VCSELs resolve this ambiguity, and, thus, the PEB decreases to approximately 4~cm, until the VCSELs exit the FoV. Moreover, for $L=2$, the opposite panel extends the localization to every orientation, while Proposition~\ref{prop1} again limits the position RMSE to 16.8~cm when the PD faces either panel. In contrast, for $L=4$, the PD always observes at least two panels from different directions, and, thus, the RSS-based estimator is employed at every orientation, with its PEB below 1~cm for 88\% of the orientations and at most 2.3~cm between two panels, where the vertical displacement of the UE and the elevation of its PD are harder to distinguish, since all the panels are mounted at the UE height. Furthermore, the RMSE closely matches its theoretical value at all the orientations, thus validating both methods. Therefore, the two methods are complementary, while millimeter-level accuracy requires the cooperation of multiple LeRIS panels.}
 
\begin{figure}
    \centering
    \begin{tikzpicture}
    \begin{axis}[
        width=0.95\linewidth,
        xlabel={\(\gamma_t\) (dB)},
        ylabel={Average data rate (Mbps)},
        xmin=60, xmax=100,
        ymin=0, ymax=700,
        xtick={60,65,...,100},
        ytick={0,100,...,700},
        grid=major,
        legend cell align={left},
        legend style={at={(0.02,0.98)}, anchor=north west, font=\footnotesize},
    ]
    \addplot[color=black, solid, line width=1pt, mark=o, mark size=2.5pt, mark repeat=5]
        table[x index=0, y index=1]{Fig4_L1_Ideal.dat};
    \addlegendentry{Ideal, L=1}
    \addplot[color=black, dashed, line width=1pt, mark=o, mark size=2.5pt, mark repeat=5, mark options={solid}]
        table[x index=0, y index=1]{Fig4_L1_Realistic.dat};
    \addlegendentry{Realistic, L=1}
    \addplot[color=blue, solid, line width=1pt, mark=triangle*, mark size=2.5pt, mark repeat=5]
        table[x index=0, y index=1]{Fig4_L2_Ideal.dat};
    \addlegendentry{Ideal, L=2}
    \addplot[color=blue, dashed, line width=1pt, mark=triangle*, mark size=2.5pt, mark repeat=5, mark options={solid}]
        table[x index=0, y index=1]{Fig4_L2_Realistic.dat};
    \addlegendentry{Realistic, L=2}
    \addplot[color=red, solid, line width=1pt, mark=star, mark size=2.5pt, mark repeat=5]
        table[x index=0, y index=1]{Fig4_L4_Ideal.dat};
    \addlegendentry{Ideal, L=4}
    \addplot[color=red, dashed, line width=1pt, mark=star, mark size=2.5pt, mark repeat=5, mark options={solid}]
        table[x index=0, y index=1]{Fig4_L4_Realistic.dat};
    \addlegendentry{Realistic, L=4}
    \end{axis}
    \end{tikzpicture}
    \caption{{Average UE data rate versus $\gamma_t$ for $L=1$, $2$, and $4$ LeRIS panels under ideal (solid) and realistic (dashed) cases.}}
    \label{fig:rate_vs_gamma}
\end{figure}

{Fig.~\ref{fig:rate_vs_gamma} presents the average UE data rate as a function of $\gamma_t$ for different numbers of participating LeRIS panels $L$, where each panel is equipped with $50\times50$ reflecting elements and $O_b=4$ blocking objects are randomly placed in the environment, while the solid and dashed curves correspond to the ideal and the realistic case, respectively. As it can be observed, involving more LeRIS panels improves the rate in both cases, since their joint participation enhances both localization and communication. In more detail, a UE whose PD observes fewer than three VCSELs cannot be localized through either Lemma~\ref{lemma1} or Proposition~\ref{prop1}, and, thus, the final LeRIS cannot be steered toward it. This frequently occurs for $L=1$, since a single panel does not provide sufficient angular diversity, i.e., a PD that faces away from it observes none of its VCSELs, whereas, for $L=4$, the panels mounted on all the walls enable the localization of most UEs. Moreover, for $L=1$, the PWE has no alternative route to circumvent the obstacles, whereas, for multiple panels, sensing enables the discovery of unobstructed alternatives, and, although the cumulative path loss increases due to the additional reflections, the resulting route diversity yields a substantially higher rate across the entire SNR range. Furthermore, the realistic case reduces the rate of every deployment, with the corresponding loss at $\gamma_t=100$~dB increasing from approximately $70$~Mbps for $L=1$ to $150$~Mbps for $L=4$, since the efficiency loss, the phase errors, and the panel-position offsets accumulate over the cascaded reflections of the multi-panel routes, while erroneous sensing decisions become more likely along routes that consist of more segments. Nevertheless, the relative gain of the four-panel system over the single-panel one remains almost unchanged in the realistic case, while, for high values of $\gamma_t$, the realistic four-panel system outperforms even the ideal single-panel one, which reveals that the additional panels compensate for the accumulated imperfections. Thus, Fig.~\ref{fig:rate_vs_gamma} highlights that the proposed VCSEL-based LeRIS architecture transforms localization and sensing into key enablers for overcoming blockages, thereby providing both reliable connectivity and significant throughput gains in obstructed indoor environments, even under practical imperfections.}
 
\begin{figure}
    \centering
    \begin{tikzpicture}
    \begin{axis}[
        width=0.95\linewidth,
        xlabel={{Number of reflecting elements}},
        ylabel={Average data rate (Mbps)},
        xmin=0, xmax=2500, 
        ymin=0, ymax=700,
        xtick={100, 500, 1000, 1500, 2000, 2500},
        xticklabel style={/pgf/number format/1000 sep={,}},
        ytick={0,100,...,700},
        grid=major,
        enlarge x limits=false, 
        legend cell align={left},
        legend style={at={(0.02,0.98)}, anchor=north west, font=\footnotesize},
    ]
    
    \addplot[color=black, solid, line width=1pt, mark=o, mark size=2.5pt] table[x index=0, y index=1]{Fig5_L1_Ideal.dat};
    \addlegendentry{Ideal, L=1}
    \addplot[color=black, dashed, line width=1pt, mark=o, mark size=2.5pt, mark options={solid}] table[x index=0, y index=1]{Fig5_L1_Realistic.dat};
    \addlegendentry{Realistic, L=1}
    
    \addplot[color=blue, solid, line width=1pt, mark=triangle*, mark size=2.5pt] table[x index=0, y index=1]{Fig5_L2_Ideal.dat};
    \addlegendentry{Ideal, L=2}
    \addplot[color=blue, dashed, line width=1pt, mark=triangle*, mark size=2.5pt, mark options={solid}] table[x index=0, y index=1]{Fig5_L2_Realistic.dat};
    \addlegendentry{Realistic, L=2}
    
    \addplot[color=red, solid, line width=1pt, mark=star, mark size=2.5pt] table[x index=0, y index=1]{Fig5_L4_Ideal.dat};
    \addlegendentry{Ideal, L=4}
    \addplot[color=red, dashed, line width=1pt, mark=star, mark size=2.5pt, mark options={solid}] table[x index=0, y index=1]{Fig5_L4_Realistic.dat};
    \addlegendentry{Realistic, L=4}
    
    \end{axis}
    \end{tikzpicture}
    \caption{{Average UE data rate versus the number of reflecting elements for $L=1$, $2$, and $4$ under ideal (solid) and realistic (dashed) cases.}}
    \label{fig:rate_vs_N}
\end{figure}
 
{Fig.~\ref{fig:rate_vs_N} presents the average UE data rate as a function of the number of reflecting elements for $L=1$, $2$, and $4$, with $\gamma_t=100$~dB. As it can be observed, increasing the number of reflecting elements improves the rate of every deployment, since it enhances the beamforming gain of each participating LeRIS. However, the localization through Lemma~\ref{lemma1} and Proposition~\ref{prop1} relies on the VCSELs, and, thus, a UE that cannot be localized remains unserved regardless of the number of reflecting elements. Therefore, for a small number of elements, where the multi-panel routes are heavily deteriorated by the path loss of the cascaded reflections, the gain of $L=4$ over $L=1$ mainly stems from the localization of more UEs. In contrast, as the number of elements grows, the increased beamforming gain counteracts this path loss and the route diversity of the additional panels is also exploited, and, thus, for $2500$ elements, $L=4$ achieves approximately twice the rate of $L=1$. This trend is also preserved in the realistic case, since the gain of each LeRIS increases quadratically with the number of reflecting elements, whereas the efficiency loss and the phase errors reduce it by an approximately constant factor, and, thus, the accumulated imperfections are increasingly compensated, so that the realistic four-panel system outperforms the ideal single-panel one already for $400$ elements. Consequently, Fig.~\ref{fig:rate_vs_N} highlights that, while $L$ determines the localization coverage and the availability of alternative routes, the number of reflecting elements determines the extent to which these routes can be exploited, thereby revealing their joint importance in scaling the performance of VCSEL-based LeRIS architectures.}

\begin{figure}
\centering
 
\begin{subfigure}[b]{\linewidth}
\centering
\begin{tikzpicture}
\begin{axis}[
    width=0.95\linewidth,
    xlabel = {\(\gamma_t\) (dB)},
    ylabel = {Average data rate (Mbps)},
    xmin = 60, xmax = 100, 
    ymin = 0, ymax = 400, 
    xtick = {60, 65, ..., 100},
    ytick = {0, 100, ..., 400},
    grid = major,
    legend style={font=\footnotesize, at={(0.05,0.95)}, anchor=north west}, 
]
\addplot[color=black, solid, mark=o, mark size=2.5pt, line width=1pt] table[x index=0, y index=1] {Fig6a_Ob2_Ideal.dat}; 
\addlegendentry{Ideal, \(O_b=2\)}
\addplot[color=black, dashed, mark=o, mark size=2.5pt, mark options={solid}, line width=1pt] table[x index=0, y index=1] {Fig6a_Ob2_Realistic.dat}; 
\addlegendentry{Realistic, \(O_b=2\)}
 
\addplot[color=blue, solid, mark=triangle*, mark size=2.5pt, line width=1pt] table[x index=0, y index=1] {Fig6a_Ob4_Ideal.dat}; 
\addlegendentry{Ideal, \(O_b=4\)}
\addplot[color=blue, dashed, mark=triangle*, mark size=2.5pt, mark options={solid}, line width=1pt] table[x index=0, y index=1] {Fig6a_Ob4_Realistic.dat}; 
\addlegendentry{Realistic, \(O_b=4\)}
 
\addplot[color=red, solid, mark=star, mark size=2.5pt, line width=1pt] table[x index=0, y index=1] {Fig6a_Ob6_Ideal.dat}; 
\addlegendentry{Ideal, \(O_b=6\)}
\addplot[color=red, dashed, mark=star, mark size=2.5pt, mark options={solid}, line width=1pt] table[x index=0, y index=1] {Fig6a_Ob6_Realistic.dat}; 
\addlegendentry{Realistic, \(O_b=6\)}
\end{axis}
\end{tikzpicture}
\caption*{(a)}
\end{subfigure}
 
\vspace{0.5cm}
 
\begin{subfigure}[b]{\linewidth}
\centering
\begin{tikzpicture}
\begin{axis}[
    width=0.95\linewidth,
    xlabel = {\(\gamma_t\) (dB)},
    ylabel = {Average data rate (Mbps)},
    xmin = 60, xmax = 100, 
    ymin = 0, ymax = 700, 
    xtick = {60, 65, ..., 100},
    ytick = {0, 100, ..., 700},
    grid = major,
    legend style={font=\footnotesize, at={(0.05,0.95)}, anchor=north west}, 
]
\addplot[color=black, solid, mark=o, mark size=2.5pt, line width=1pt] table[x index=0, y index=1] {Fig6b_Ob2_Ideal.dat}; 
\addlegendentry{Ideal, \(O_b=2\)}
\addplot[color=black, dashed, mark=o, mark size=2.5pt, mark options={solid}, line width=1pt] table[x index=0, y index=1] {Fig6b_Ob2_Realistic.dat}; 
\addlegendentry{Realistic, \(O_b=2\)}
 
\addplot[color=blue, solid, mark=triangle*, mark size=2.5pt, line width=1pt] table[x index=0, y index=1] {Fig6b_Ob4_Ideal.dat}; 
\addlegendentry{Ideal, \(O_b=4\)}
\addplot[color=blue, dashed, mark=triangle*, mark size=2.5pt, mark options={solid}, line width=1pt] table[x index=0, y index=1] {Fig6b_Ob4_Realistic.dat}; 
\addlegendentry{Realistic, \(O_b=4\)}
 
\addplot[color=red, solid, mark=star, mark size=2.5pt, line width=1pt] table[x index=0, y index=1] {Fig6b_Ob6_Ideal.dat}; 
\addlegendentry{Ideal, \(O_b=6\)}
\addplot[color=red, dashed, mark=star, mark size=2.5pt, mark options={solid}, line width=1pt] table[x index=0, y index=1] {Fig6b_Ob6_Realistic.dat}; 
\addlegendentry{Realistic, \(O_b=6\)}
\end{axis}
\end{tikzpicture}
\caption*{(b)}
\end{subfigure}
 
\caption{{Average UE data rate versus $\gamma_t$ for different obstacle counts $O_b$ under ideal (solid) and realistic (dashed) cases: (a) $L=1$ and (b) $L=4$.}}
\label{fig:vertical_ris_bandwidth}
\end{figure}
 
{Fig.~\ref{fig:vertical_ris_bandwidth} presents the average UE data rate as a function of $\gamma_t$ for $O_b\in\{2,4,6\}$ obstacles, for (a) $L=1$ and (b) $L=4$. As it can be observed, additional obstacles reduce the rate of both deployments, since they block not only the route segments, but also the optical LoS of the VCSELs, and, thus, reduce the VCSELs available for localization. This effect is more pronounced for $L=1$, since the VCSELs of a single panel are easily obstructed, while the absence of alternative routes leaves the UE unserved whenever its route is blocked. In contrast, for $L=4$, the panels on all the walls enable the localization of most UEs even for $O_b=6$, while sensing identifies unobstructed alternative routes, and, thus, the relative rate reduction is lower. However, in the realistic case, this robustness is partially reduced, since more obstructed segments increase the probability that a route is erroneously identified as feasible, which yields zero rate, while multi-panel routes consist of more segments that may be erroneously identified as blocked. Nevertheless, for every $O_b$, the four-panel system achieves approximately twice the rate of the single-panel one in both cases, while, for high values of $\gamma_t$, the realistic four-panel system with $O_b=6$ outperforms even the ideal single-panel one with $O_b=2$. Therefore, Fig.~\ref{fig:vertical_ris_bandwidth} highlights that multiple LeRIS panels enhance the robustness of both localization and communication against blockages, even in heavily obstructed environments.}

\begin{figure}
  \centering
  
  \begin{subfigure}[b]{\linewidth}
    \centering
    \begin{tikzpicture}
      \begin{axis}[
        width=0.95\linewidth,
        xlabel = {Minimum user data rate (Mbps)},
        ylabel = {CDF},
        xmin = 0, xmax = 1600,
        xtick = {0,  400, 800, 1200, 1600},
        ymin = 0, ymax = 1,
        ytick = {0,  0.2, 0.4,  0.6, 0.8,  1},
        grid = major,
        legend style={font=\scriptsize, at={(0.98,0.02)}, anchor=south east},
      ]
        \addplot[color=black, solid, line width=1pt] table {Fig7a_L1_Ideal.dat};
        \addlegendentry{Ideal, K=1, L=1}
        \addplot[color=black, dashed, line width=1pt] table {Fig7a_L1_Realistic.dat};
        \addlegendentry{Realistic, K=1, L=1}
        
        \addplot[color=blue, solid, line width=1pt] table {Fig7a_L2_Ideal.dat};
        \addlegendentry{Ideal, K=1, L=2}
        \addplot[color=blue, dashed, line width=1pt] table {Fig7a_L2_Realistic.dat};
        \addlegendentry{Realistic, K=1, L=2}
        
        \addplot[color=red, solid, line width=1pt] table {Fig7a_L4_Ideal.dat};
        \addlegendentry{Ideal, K=1, L=4}
        \addplot[color=red, dashed, line width=1pt] table {Fig7a_L4_Realistic.dat};
        \addlegendentry{Realistic, K=1, L=4}
      \end{axis}
    \end{tikzpicture}
    \caption{}
  \end{subfigure}
  
  \vspace{0.5cm}
  
  \begin{subfigure}[b]{\linewidth}
    \centering
    \begin{tikzpicture}
      \begin{axis}[
        width=0.95\linewidth,
        xlabel = {Minimum user data rate (Mbps)},
        ylabel = {CDF},
        xmin = 0, xmax = 250,
        xtick = {0,  50,  100,  150,   200, 250},
        ymin = 0.5, ymax = 1,
        ytick = {0.5, 0.6, 0.7, 0.8, 0.9, 1},
        grid = major,
        legend style={font=\scriptsize, at={(0.98,0.02)}, anchor=south east},
      ]
        \addplot[color=black, solid, line width=1pt] table {Fig7b_L1_Ideal.dat};
        \addlegendentry{Ideal, K=5, L=1}
        \addplot[color=black, dashed, line width=1pt] table {Fig7b_L1_Realistic.dat};
        \addlegendentry{Realistic, K=5, L=1}
        
        \addplot[color=blue, solid, line width=1pt] table {Fig7b_L2_Ideal.dat};
        \addlegendentry{Ideal, K=5, L=2}
        \addplot[color=blue, dashed, line width=1pt] table {Fig7b_L2_Realistic.dat};
        \addlegendentry{Realistic, K=5, L=2}
        
        \addplot[color=red, solid, line width=1pt] table {Fig7b_L4_Ideal.dat};
        \addlegendentry{Ideal, K=5, L=4}
        \addplot[color=red, dashed, line width=1pt] table {Fig7b_L4_Realistic.dat};
        \addlegendentry{Realistic, K=5, L=4}
      \end{axis}
    \end{tikzpicture}
    \caption{}
  \end{subfigure}
  
  \caption{{CDF of the minimum user data rate for (a) $K=1$ and (b) $K=5$ under ideal (solid) and realistic (dashed) cases.}}
  \label{fig:CDF vs R}
\end{figure}
 
{Finally, Fig.~\ref{fig:CDF vs R} presents the cumulative distribution function of the minimum user data rate, which is the outcome of the TDMA-based max--min optimization problem in \eqref{eq:TDMA}, for $\gamma_t=100$~dB. In the single-user case, the entire transmission time is devoted to the UE, and, thus, the CDF at zero rate corresponds to the outage probability, which arises when the UE cannot be localized, since its PD observes fewer than three VCSELs, or when no unobstructed route is available. As it can be observed, increasing the number of LeRIS panels reduces the outage probability from approximately $0.66$ for $L=1$ to $0.23$ for $L=4$, while it also shifts the distribution to higher rates, since the additional panels enable both the localization of more UEs and the discovery of unobstructed alternative routes. In the multi-user case with $K=5$, the minimum rate is zero whenever any of the UEs is not served, and, thus, the outage probability increases for every deployment, while the max--min allocation assigns more transmission time to the UEs with lower rates, which limits the minimum rate to approximately $200$~Mbps. In more detail, in the ideal case, all five UEs are served in only $1\%$ of the realizations for $L=1$, whereas, for $L=4$, this percentage increases to approximately $38\%$, which reveals that the benefit of multiple LeRIS panels becomes more pronounced as the number of UEs increases. Moreover, the realistic case increases the outage probability of every deployment, since an erroneous sensing decision for any of the UEs may nullify the minimum rate, while the remaining imperfections shift the curves to lower rates. Nevertheless, in both subfigures, the realistic four-panel system achieves a lower outage probability than even the ideal two-panel one, which reveals that the route diversity of multiple LeRIS panels outweighs the impact of the imperfect blockage information. Therefore, Fig.~\ref{fig:CDF vs R} demonstrates that, while, in the single-user case, multiple LeRIS panels primarily enhance reliability by mitigating outages, in the multi-user case they are essential for fairness, as only their cooperation enables the simultaneous localization and service of all UEs.}

\section{Conclusion}
In this work, a VCSEL-based LeRIS architecture was developed to jointly support user localization, environmental mapping, and mmWave communication in PWEs. By exploiting the narrow Gaussian beams and {the beam scanning} of VCSELs, closed-form schemes were obtained for the joint recovery of the user position and orientation, showing that five VCSELs are sufficient {from their RSS values, while only three VCSELs are required when the pointing directions of their beams are exploited} under the specified geometric conditions. Furthermore, the corresponding PEB was derived, while reflected-signal time-of-arrival measurements were employed to identify obstructed links and enable blockage-aware LeRIS route selection. The obtained results revealed that {the cooperation of multiple LeRIS panels is essential for millimeter-level localization, since it resolves the ambiguity between the displacement of the UE and the tilt of its PD that arises when a single panel is observed, and, thus, maintains accurate} localization independently of the UE orientation. Moreover, the resulting route diversity was shown to mitigate blockage effects and improve both {the average rate and the minimum user rate}{, even under non-unit reflection efficiency, hardware imperfections in the RIS configuration, panel-position offsets, and erroneous sensing decisions, since the additional routes compensate for the imperfections that accumulate over the cascaded reflections}. Therefore, the proposed framework establishes how compact optical sources can extend LeRIS operation beyond localization and provide the spatial awareness required for resilient and adaptive PWEs.
 
\appendices
\section{Levenberg–Marquardt Method for Localization Through VCSELs} \label{App1}
The minimization problem in \eqref{minproblem} is solved using the Levenberg–Marquardt (LM) method, whose objective is to iteratively refine the position estimate $\mathbf{r}$ \cite{LM2003}. This requires, at each step, the evaluation of the residual vector and the Jacobian with respect to $\mathbf{r}$, which together determine the update direction. In this context, the residual vector for the examined VCSEL-based localization task is defined as
\begin{small}
\begin{equation}
\mathbf{e}(\mathbf{r}) = V(\mathbf{r})\hat{\mathbf{n}}(\mathbf{r}) - \mathbf{p},
\end{equation}    
\end{small}
{where $\mathbf{p}=[P_1,\ldots,P_{N_{\mathrm{V}}}]^T$ denotes the received powers from the $N_{\mathrm{V}}$ visible VCSELs, while the $i$-th entry of $\mathbf{e}(\mathbf{r})$ represents the mismatch between the modeled and observed RSS, i.e.,
\begin{small}
 \begin{equation}
f_i(\mathbf{r},\mathbf{n})=\beta_i(d_i)\,\mathbf{n}^T\mathbf{u}_i-P_i.
\end{equation}   
\end{small}
}To obtain the update step, the Jacobian of the residuals with respect to $\mathbf{r}$ must be derived, which can be obtained by differentiating $f_i(\mathbf{r},\mathbf{n})$ while holding $\mathbf{n}=\hat{\mathbf{n}}(\mathbf{r})$ fixed, yielding 
\begin{small}
\begin{equation}
\nabla_{\mathbf r} f_i = \beta_i'(d_i)({\mathbf{n}^T\mathbf{u}_i})\,\nabla_{\mathbf r} d_i + \beta_i(d_i)\,\nabla_{\mathbf r}({\mathbf{n}^T\mathbf{u}_i}).
\label{nabla1}
\end{equation}    
\end{small}Thus, by noting that $\frac{\partial d_i}{\partial \mathbf r}=-\mathbf u_i$, and $
\nabla_{\mathbf r}({\mathbf{n}^T\mathbf{u}_i})= -\frac{1}{d_i}(I-\mathbf u_i\mathbf u_i^{T})\mathbf n$, we can substitute these equations in \eqref{nabla1}, which then can be rewritten as
\begin{small}
 \begin{equation}
\nabla_{\mathbf{r}} f_i
= -\beta_i'(d_i)({\mathbf{n}^T\mathbf{u}_i})\mathbf{u}_i
-\frac{\beta_i(d_i)}{d_i}\big(I-\mathbf{u}_i\mathbf{u}_i^{T}\big)\mathbf{n}.
\label{eq:grad_general_appendix}
\end{equation}   
\end{small}Moreover, considering that the UE is located in the far field of the VCSELs, it holds that $\beta_i(d)=K_i/d^2$ and $\beta_i'(d)=-2\beta_i(d)/d$, hence, by substituting these relations into
\eqref{eq:grad_general_appendix}, we obtain
\begin{small}
\begin{equation}
\nabla_{\mathbf r} f_i
= \frac{\beta_i(d_i)}{d_i}\Big(2({\mathbf n^T\mathbf u_i})\,\mathbf u_i
-\big(I-\mathbf u_i\mathbf u_i^{T}\big)\mathbf n\Big).
\label{eq:grad_farfield_appendix}
\end{equation}    
\end{small}The gradients of all residuals are then collected into the Jacobian matrix $G_{J}(\mathbf{r},\mathbf{n})\in\mathbb{R}^{{N_{\mathrm{V}}}\times 3}$, while the residual vector can be written as $\mathbf{e}(\mathbf{r})=[f_1(\mathbf{r},\mathbf{n}),\dots,f_{{N_{\mathrm{V}}}}(\mathbf{r},\mathbf{n})]^{T}$. With these quantities available, the LM update step is obtained as the solution of
\begin{small}
 \begin{equation}
(G_{J}^{T} G_{J}+\lambda_d I_3)\,\Delta\mathbf{r}=-G^{T}_{J} \mathbf{e},
\label{eq:LM_step_appendix}
\end{equation}   
\end{small}where $\lambda_d>0$ is the damping parameter. {It should be noted that $G_J$ is the partial Jacobian of $\mathbf{e}(\mathbf{r})$, since it does not include the term $V(\mathbf{r})\,\partial\hat{\mathbf{n}}/\partial\mathbf{r}^T$, where $\hat{\mathbf{n}}(\mathbf{r})$ is differentiable with respect to $\mathbf{r}$ by the implicit function theorem, since $\phi'(\lambda)<0$ at $\hat{\lambda}$, which exists with probability one for noisy RSS values. However, the contribution $(\partial\hat{\mathbf{n}}/\partial\mathbf{r}^T)^TV^T\mathbf{e}$ of this term to the gradient vanishes, because $V^T\mathbf{e}=-\hat{\lambda}\hat{\mathbf{n}}$ from \eqref{eq:reg_normal} and $\|\hat{\mathbf{n}}(\mathbf{r})\|_2=1$ for every $\mathbf{r}$, and, thus, $G_J^T\mathbf{e}$ in \eqref{eq:LM_step_appendix} is the exact gradient of $J(\mathbf{r})$, in direct analogy with the simplified Jacobian of the variable-projection method \cite{LM2003}.} {Nevertheless, since the omitted term affects only the curvature, the complete Jacobian $G_J+V(\mathbf{r})\,\partial\hat{\mathbf{n}}/\partial\mathbf{r}^T$, where $\partial\hat{\mathbf{n}}/\partial\mathbf{r}^T$ follows by differentiating \eqref{eq:reg_normal} under the constraint $\|\hat{\mathbf{n}}(\mathbf{r})\|_2=1$, can also be employed in \eqref{eq:LM_step_appendix} without altering the gradient and, thus, the stationary points, which accelerates the convergence when the position and the orientation of the UE are strongly coupled, e.g., when a single LeRIS panel is observed, and, hence, it is adopted in Section~\ref{sec:IV}.} Finally, the UE position estimate is updated as $\mathbf{r}\leftarrow \mathbf{r}+\Delta\mathbf{r}$ whenever the actual decrease in $J(\mathbf{r})$ agrees sufficiently with the decrease predicted by the local quadratic model, while $\lambda_d$ is adjusted at each iteration to balance stability and convergence. Thus, the LM process refines $\mathbf{r}$ until convergence, with $\mathbf{n}=\hat{\mathbf{n}}(\mathbf{r})$ always determined by the separable structure of the problem. 
 
\section{Proof of Proposition~\ref{prop:joint_fim_peb}}
\label{app:joint_fim_proof} 
To prove Proposition~\ref{prop:joint_fim_peb}, we first derive
the gradient of each mean RSS observation with respect to the
joint position-and-orientation state of UE $k$. We then use
the Gaussian observation model to obtain the Fisher
information contributed by each visible VCSEL and combine
these contributions to derive the joint FIM and the resulting
PEB. For compactness, we define
\begin{small}
\begin{equation}
c_{i,k}
\triangleq
\mathbf n_k^{T}\mathbf u_{i,k},
\label{eq:appendix_incidence_factor}
\end{equation}    
\end{small}such that the mean RSS in \eqref{eq:fim_mean_rss} can be
written as
$\mu_{i,k}=\beta_i(d_{i,k})c_{i,k}$.
 
We first derive the position component of
$\mathbf g_{i,k}$. From the definitions of $d_{i,k}$ and
$\mathbf u_{i,k}$, it follows that
\begin{small}
 \begin{equation}
\nabla_{\mathbf r_k}d_{i,k}
=
-\mathbf u_{i,k},
\label{eq:appendix_distance_gradient}
\end{equation}   
\end{small}and
\begin{small}
 \begin{equation}
\frac{\partial\mathbf u_{i,k}}
{\partial\mathbf r_k^{T}}
=
-\frac{1}{d_{i,k}}
\left(
\mathbf I_3
-
\mathbf u_{i,k}\mathbf u_{i,k}^{T}
\right).
\label{eq:appendix_direction_jacobian}
\end{equation}   
\end{small}Therefore,
\begin{small}
 \begin{equation}
\nabla_{\mathbf r_k}c_{i,k}
=
-\frac{1}{d_{i,k}}
\left(
\mathbf I_3
-
\mathbf u_{i,k}\mathbf u_{i,k}^{T}
\right)
\mathbf n_k.
\label{eq:appendix_incidence_gradient}
\end{equation}   
\end{small}Moreover, differentiating $\beta_i(d)$ with respect to $d$
gives
\begin{small}
\begin{equation}
\beta_i'(d)
=
-\beta_i(d)
\frac{2d}{z_{R,i}^{2}+d^{2}}.
\label{eq:appendix_beta_derivative}
\end{equation}    
\end{small} and by applying the product and chain rules yields
\begin{small}
\begin{align}
\mathbf g_{r,i,k}
&\triangleq
\nabla_{\mathbf r_k}\mu_{i,k}
\nonumber\\
&=
\beta_i'(d_{i,k})c_{i,k}
\nabla_{\mathbf r_k}d_{i,k}
+
\beta_i(d_{i,k})
\nabla_{\mathbf r_k}c_{i,k}
\nonumber\\
&=
-\beta_i'(d_{i,k})c_{i,k}\mathbf u_{i,k}
-
\frac{\beta_i(d_{i,k})}{d_{i,k}}
\left(
\mathbf I_3
-
\mathbf u_{i,k}\mathbf u_{i,k}^{T}
\right)
\mathbf n_k.
\label{eq:appendix_position_gradient}
\end{align}   
\end{small}

We next obtain the orientation components of the gradient.
Using the PD orientation vector in
\eqref{eq:pd_orientation_vector}, and evaluating the angular
derivatives in radians, we have
\begin{small}
\begin{equation}
\mathbf n_{\varphi,k}
\triangleq
\frac{\partial\mathbf n_k}{\partial\varphi_k}
=
\begin{bmatrix}
-\cos\vartheta_k\sin\varphi_k\\
\cos\vartheta_k\cos\varphi_k\\
0
\end{bmatrix},
\label{eq:appendix_azimuth_normal_derivative}
\end{equation}    
\end{small}and
\begin{small}
\begin{equation}
\mathbf n_{\vartheta,k}
\triangleq
\frac{\partial\mathbf n_k}{\partial\vartheta_k}
=
\begin{bmatrix}
-\sin\vartheta_k\cos\varphi_k\\
-\sin\vartheta_k\sin\varphi_k\\
\cos\vartheta_k
\end{bmatrix}.
\label{eq:appendix_elevation_normal_derivative}
\end{equation}   
\end{small}Since $\beta_i(d_{i,k})$ does not depend on the orientation
angles, it follows that
\begin{small}
\begin{equation}
g_{\varphi,i,k}
\triangleq
\frac{\partial\mu_{i,k}}{\partial\varphi_k}
=
\beta_i(d_{i,k})
\mathbf n_{\varphi,k}^{T}\mathbf u_{i,k},
\label{eq:appendix_azimuth_gradient}
\end{equation}    
\end{small}and
\begin{small}
\begin{equation}
g_{\vartheta,i,k}
\triangleq
\frac{\partial\mu_{i,k}}{\partial\vartheta_k}
=
\beta_i(d_{i,k})
\mathbf n_{\vartheta,k}^{T}\mathbf u_{i,k}.
\label{eq:appendix_elevation_gradient}
\end{equation}    
\end{small}Thus, the complete measurement gradient in
\eqref{eq:fim_measurement_gradient} is
\begin{small}
 \begin{equation}
\mathbf g_{i,k}
=
\begin{bmatrix}
\mathbf g_{r,i,k}\\
g_{\varphi,i,k}\\
g_{\vartheta,i,k}
\end{bmatrix}.
\label{eq:appendix_complete_gradient}
\end{equation}   
\end{small}

Under the Gaussian observation model in \eqref{eq:fim_rss_observation}, with the equivalent optical-noise variance defined in
\eqref{eq:equivalent_optical_noise_variance}, the log-likelihood contribution of the
$i$-th observation is
\begin{small}
\begin{equation}
{\Lambda_{i,k}}
=
{-\frac{1}{2}\ln\left(2\pi P_{n,i,k}^{2}\right)}
-\frac{1}{2P_{n,i,k}^{2}}
\left[
P_{r,i,k}
-
\mu_{i,k}(\boldsymbol{\xi}_k)
\right]^2{,}
\label{eq:appendix_log_likelihood}
\end{equation}    
\end{small}

{where, according to \eqref{eq:equivalent_optical_noise_variance} and \eqref{eq:electrical_noise_psd}, the noise variance depends on $\boldsymbol{\xi}_k$ exclusively through $\mu_{i,k}$, since $P_{\mathrm{LoS},i,k}=\mu_{i,k}$, and, thus, $\nabla_{\boldsymbol{\xi}_k}P_{n,i,k}^{2}=\varsigma_{i,k}\mathbf g_{i,k}$, with $\varsigma_{i,k}=\partial P_{n,i,k}^{2}/\partial\mu_{i,k}=2B_o\left(q/R_{\mathrm{PD}}+\mathrm{RIN}\,\mu_{i,k}\right)$.}
Differentiating \eqref{eq:appendix_log_likelihood} with
respect to the unknown UE state yields
\begin{small}
\begin{equation}
\begin{split}
\nabla_{\boldsymbol{\xi}_k}\Lambda_{i,k}
&=
\left[
\frac{
P_{r,i,k}
-
\mu_{i,k}(\boldsymbol{\xi}_k)
}{
P_{n,i,k}^{2}
}
\right. \\
&\qquad
\left.
+\varsigma_{i,k}
\frac{\left(P_{r,i,k}-\mu_{i,k}(\boldsymbol{\xi}_k)\right)^{2}-P_{n,i,k}^{2}}
{2P_{n,i,k}^{4}}
\right]\mathbf g_{i,k}.
\end{split}
\label{eq:appendix_log_likelihood_gradient}
\end{equation}
\end{small}Since {$P_{r,i,k}-\mu_{i,k}(\boldsymbol{\xi}_k)$ is zero-mean Gaussian with variance $P_{n,i,k}^{2}$, its third central moment is zero, while}
\begin{small}
\begin{equation}
\mathbb E
\left[
\left(
\left(P_{r,i,k}-\mu_{i,k}(\boldsymbol{\xi}_k)\right)^2
-P_{n,i,k}^{2}
\right)^2
\right]
=
2P_{n,i,k}^{4},
\label{eq:appendix_residual_variance}
\end{equation}    
\end{small}{and, thus,} the Fisher information contributed by this observation is {\cite{Kay1993}}
\begin{small}
\begin{align}
\mathbf J_{i,k}
&=
\mathbb E
\left[
\left(
\nabla_{\boldsymbol{\xi}_k}{\Lambda_{i,k}}
\right)
\left(
\nabla_{\boldsymbol{\xi}_k}{\Lambda_{i,k}}
\right)^{T}
\right]
\nonumber\\
&=
\frac{{1+\varepsilon_{i,k}}}{P_{n,i,k}^{2}}
\mathbf g_{i,k}\mathbf g_{i,k}^{T}.
\label{eq:appendix_single_measurement_fim}
\end{align}    
\end{small}Furthermore, since the VCSEL observations are statistically independent,
their Fisher-information contributions are additive.
Therefore,
\begin{small}
\begin{equation}
\mathbf J_k^{(L)}
=
\sum_{i\in\mathcal V_k^{(L)}}
\frac{{1+\varepsilon_{i,k}}}{P_{n,i,k}^{2}}
\mathbf g_{i,k}\mathbf g_{i,k}^{T},
\label{eq:appendix_joint_fim}
\end{equation}    
\end{small}which establishes \eqref{eq:joint_state_fim}. For any unbiased estimator of the complete UE state, the
matrix Cramér--Rao lower bound (CRLB) gives
\begin{small}
\begin{equation}
\operatorname{Cov}
\left(
\widehat{\boldsymbol{\xi}}_k
\right)
\succeq
\left(
\mathbf J_k^{(L)}
\right)^{-1}.
\label{eq:appendix_joint_crlb}
\end{equation}    
\end{small}Since the first three elements of $\boldsymbol{\xi}_k$
correspond to the UE position, the associated position
covariance satisfies
\begin{small}
\begin{equation}
\operatorname{Cov}
\left(
\widehat{\mathbf r}_k
\right)
\succeq
\left[
\left(
\mathbf J_k^{(L)}
\right)^{-1}
\right]_{1:3,1:3}.
\label{eq:appendix_position_crlb}
\end{equation}    
\end{small}Consequently,
\begin{small}
\begin{align}
\mathbb E
\left[
\left\|
\widehat{\mathbf r}_k-\mathbf r_k
\right\|_2^2
\right]
&=
\operatorname{tr}
\left[
\operatorname{Cov}
\left(
\widehat{\mathbf r}_k
\right)
\right]
\nonumber\\
&\geq
\operatorname{tr}
\left(
\left[
\left(
\mathbf J_k^{(L)}
\right)^{-1}
\right]_{1:3,1:3}
\right).
\label{eq:appendix_position_mse_bound}
\end{align}    
\end{small}Finally, by taking the square root of both sides yields
\begin{small}
\begin{equation}
\sqrt{
\mathbb E
\left[
\left\|
\widehat{\mathbf r}_k-\mathbf r_k
\right\|_2^2
\right]
}
\geq
\sqrt{
\operatorname{tr}
\left(
\left[
\left(
\mathbf J_k^{(L)}
\right)^{-1}
\right]_{1:3,1:3}
\right)
}
=
\mathrm{PEB}_k^{(L)},
\label{eq:appendix_peb_result}
\end{equation}    
\end{small}

which establishes \eqref{eq:position_error_bound} and
completes the proof.
 
\bibliographystyle{IEEEtran}
\bibliography{Bibliography_1.bib}
\end{document}